\documentclass[12pt]{article}
\pdfoutput=1
\usepackage{amsmath}
\usepackage{times}
\usepackage{graphicx}
\usepackage{color}
\usepackage{multirow}
\usepackage[section]{placeins}
\usepackage{hyperref}
\usepackage{mathtools}
\usepackage{siunitx}
\usepackage{adjustbox}

\usepackage[authoryear]{natbib}
\usepackage{rotating}
\usepackage{bbm}
\usepackage{latexsym}
\newcommand{\captionfonts}{\normalsize}

\makeatletter  
\long\def\@makecaption#1#2{%
  \vskip\abovecaptionskip
  \sbox\@tempboxa{{\captionfonts #1: #2}}%
  \ifdim \wd\@tempboxa >\hsize
    {\captionfonts #1: #2\par}
  \else
    \hbox to\hsize{\hfil\box\@tempboxa\hfil}%
  \fi
  \vskip\belowcaptionskip}
\makeatother   
\DeclareMathOperator*{\argmin}{arg\,min}

\makeatletter
\newcounter{subsubparagraph}[subparagraph]
\renewcommand\thesubsubparagraph{%
  \thesubparagraph.\@arabic\c@subsubparagraph}
\newcommand\subsubparagraph{%
  \@startsection{subsubparagraph}    % counter
    {6}                              % level
    {\parindent}                     % indent
    {3.25ex \@plus 1ex \@minus .2ex} % beforeskip
    {-1em}                           % afterskip
    {\normalfont\normalsize\bfseries}}
\newcommand\l@subsubparagraph{\@dottedtocline{6}{10em}{5em}}
\newcommand{\subsubparagraphmark}[1]{}
\makeatother
\begin{document}

%\title{differential covariance: A New Method to Estimate Sparse Connectivity from Neural Recordings}
\hspace{13.9cm}1

\ \vspace{20mm}\\

{\LARGE Differential Covariance: A New Class of \mbox{Methods} to Estimate Sparse Connectivity from Neural \mbox{Recordings}}

\ \\
{\bf \large Tiger W. Lin$^{\displaystyle 1, \displaystyle 2}$, Anup Das$^{\displaystyle 1, \displaystyle 5}$, Giri P. Krishnan$^{\displaystyle 4}$, Maxim Bazhenov$^{\displaystyle 4}$, Terrence J. Sejnowski$^{\displaystyle 1, \displaystyle 3}$}\\
{$^{\displaystyle 1}$Howard Hughes Medical Institute, Computational Neurobiology Laboratory, Salk Institute for Biological Studies, La Jolla, CA 92037.}\\
{$^{\displaystyle 2}$Neurosciences Graduate Program, University of California San Diego, La Jolla, CA 92092.}\\
{$^{\displaystyle 3}$Institute for Neural Computation, University of California San Diego, La Jolla, CA 92092.}\\
{$^{\displaystyle 4}$Department of Medicine, University of California San Diego, La Jolla, CA 92092.}\\
{$^{\displaystyle 5}$Jacobs School of Engineering, University of California San Diego, La Jolla, CA 92092.}\\
%

%\ \\[-2mm]

\thispagestyle{empty}
\markboth{}{NC instructions}
\ \vspace{-0mm}\\
%\author[1,2]{Tiger W. Lin}
%\author[1,5]{Anup Das}
%\author[4]{Giri P. Krishnan}
%\author[4]{Maxim Bazhenov}
%\author[1,3]{Terrence J. Sejnowski}
%\affil[1]{Howard Hughes Medical Institute, Computational Neurobiology Laboratory, Salk Institute for Biological Studies, La Jolla, CA 92037}
%\affil[2]{Neurosciences Graduate Program, University of California San Diego, La Jolla, CA 92092}
%\affil[3]{Institute for Neural Computation, University of California San Diego, La Jolla, CA 92037}
%\affil[4]{Department of Cell Biology and Neuroscience, University of California, Riverside, CA 92521}
%\affil[5]{Jacobs School of Engineering, University of California San Diego, La Jolla, CA 92092}
%\renewcommand\Authands{ and }

%\date{}
%\maketitle
%\begin{abstract}
%Abstract
\begin{center} {\bf Abstract} \end{center}
With our ability to record more neurons simultaneously, making sense of these data is a challenge.
Functional connectivity is one popular way to study the relationship between multiple neural signals.
Correlation-based methods are a set of currently well-used  techniques for functional connectivity estimation. However, due to explaining away and unobserved common inputs \citep{stevenson2008inferring}, they produce spurious connections. 
The general linear model (GLM), which models spikes trains as Poisson processes \citep{okatan2005analyzing,truccolo2005point, pillow2008spatio}, avoids these confounds.
We develop here a new class of methods by using differential signals based on simulated intracellular voltage recordings. 
It is equivalent to a regularized AR(2) model.
We also expand the method to simulated local field potential (LFP) recordings and calcium imaging. 
In all of our simulated data, the differential covariance-based methods achieved better or similar performance to the GLM method and required fewer data samples. This new class of methods provides alternative ways to analyze neural signals.
%\end{abstract}

\section{Introduction}
\label{sec_intro}
%\subsection{The problem, the significance to estimate the true wiring diagram}
Simultaneous recording of large population of neurons is an inexorable trend in current neuroscience research \citep{kandel2013neuroscience}. 
Over the last five decades, the number of simultaneously recorded neurons doubles approximately every 7 years \citep{stevenson2011advances}. 
One way to make sense of this big data is to measure the functional connectivity between neurons \citep{friston2011functional}, and link the function of the neural circuit to behavior.
%\subsection{the current status quo}
As previously reviewed \citep{stevenson2008inferring}, correlation-based methods have been used to estimate functional connectivity for a long time. However, they are suffering from the problem of explaining away and unobserved common inputs \citep{stevenson2008inferring}, which make it difficult to interpret the link between the estimated correlation and the physiological network. 
More recently, \cite{okatan2005analyzing,truccolo2005point, pillow2008spatio} applied the generalized linear model to spike train data and showed good probabilistic modeling of the data.

%\subsection{we developed a new method which has these advantages}

To overcome these issues, we developed a new class of methods that use not only the spike trains but the voltages of the neurons. They achieve better performance to the GLM method but are free from the Poisson process model and require fewer data samples. They provide directionality information about sources and sinks in a network, which is important to determine the hierarchical structure of a neural circuit. 

In this paper, we further show that our differential covariance method is equivalent to a regularized second-order multivariate autoregressive model. Multivariate autoregressive (MAR) model has been used before to analyze neuroscience data \citep{friston2011functional, mcintosh1991structural}. In particular, MAR model with an arbitrary order has been discussed before \citep{harrison2003multivariate}. Continuous AR process, which is known as  Ornstein-Uhlenbeck (OU) process \citep{uhlenbeck1930theory}, has also been applied to model neuronal activity \citep{burkitt2006review, ricciardi1979ornstein, lansky1995ornstein}.
However, the modified AR(2) model in this paper haven't been used before.

%\subsection{we tested the method with a reasonably realistic model}

In this paper, all of the data that we analyze are simulated data so that we can compare different methods with ground truth. 
We first generated data using a simple passive neuron model, and provide theoretical proof for why the new methods perform better.
Then, we used a more realistic Hodgkin-Huxley (HH) based thalamocortical model to simulate intracellular recordings and local field potential data. 
%Prior works \citep{bazhenov02-8691, chen12-3987, bonjean11-9124} have demonstrated that 
This model can successfully generate sleep patterns such as spindles and slow waves \citep{bazhenov02-8691, chen12-3987, bonjean11-9124}. Since the model has a cortical layer and a thalamic layer, we further assume that the neural signals in the cortical layer are visible by the recording instruments while those from the thalamic layer are not. This is a reasonable assumption for many experiments. Since, thalamus is a deep brain structure, most experiments  involve only measurements from cortex.

Next, we simulated 1000 Hodgkin-Huxley neurons networks with 80\% excitatory neurons and 20\% inhibitory neurons sparsely connected. As in real experiments, We recorded simulated calcium signals from only a small percentage of the network (50 neurons) and compared the performance of different methods. In all simulations, our differential covariance-based methods achieve better or similar performance to the GLM method. And in the LFP and calcium imaging dataset, they achieve the same performance with fewer data samples.

The paper is organized as follow: In section~\ref{sec_methods}, we introduce our new methods. In section~\ref{sec_results}, we show the performance of all methods and explain why our methods perform better. In section~\ref{sec_discussion}, we discuss the advantage and generalizability of our methods. We also propose a few improvements that can be done in the future.

\section{Methods}
\label{sec_methods}
\subsection{Simulation models used to benchmark the methods}
\subsubsection{Passive neuron model}
To validate and test our new methods, we first developed a passive neuron model. Because of its simplicity, we can provide some theoretical proof for why our new class of methods are better. Every neuron in this model has a passive cell body with capacitance $C$ and a leakage channel with conductance $g_l$. Neurons are connected with a directional synaptic conductance $g_{syn}$; For example, neuron $i$ receiving inputs from neurons $i-4$ and $i-3$:
\begin{equation}
C\frac{dV_i}{dt}=g_{syn}V_{i-4} + g_{syn}V_{i-3} + g_{l}V_i+\mathcal{N}_i
\end{equation}
Here, we let $C=1,g_{syn}=3,g_{l}=-5$, and $\mathcal{N}_i$ is a Gaussian noise with standard deviation of 1. The connection pattern is shown in Fig.\ref{circuitOld}A. There are 60 neurons in this circuit. The first 50 neurons are connected with a pattern that: neuron $i$ projects to neuron $i+3$ and $i+4$. To make the case more realistic, aside from these 50 neurons that are visible, we added 10 more neurons (neuron 51 to 60 in Fig.\ref{circuitOld}A) that are invisible during our estimations (i.e. only the membrane voltages of the first 50 neurons are used to estimate connectivity). These 10 neurons send latent inputs to the visible neurons and introduce external correlations into the system. Therefore, we update our passive neuron's model as:
\begin{equation}
C\frac{dV_i}{dt}=g_{syn}V_{i-4} + g_{syn}V_{i-3} + g_{latent}V_{latent1} + g_{l}V_i+\mathcal{N}_i
\label{passiveModel}
\end{equation}
where $g_{latent}=10$. We choose the latent input's strength in the same scale as other connections in the network. We tested multiple values between [0, 10], higher value generates more interference and therefore makes the case more difficult.

We added the latent inputs to the system because unobserved common inputs exist in real-world problems \citep{stevenson2008inferring}.
For example, one could be using two-photon imaging to record calcium signals from the cortical circuit. The cortical circuit might receive synaptic inputs from deeper layers in the brain, such as the thalamus, which is not visible to the microscopy. Each invisible neuron projects to many visible neurons leading to common synaptic currents to the cortical circuit and causes neurons in the cortical circuit to be highly correlated. 
Later, we discuss how to remove interference from the latent inputs.
\subsubsection{Thalamocortical model}
\begin{figure}[hbtp]
\begin{center}

\includegraphics[width=\textwidth]{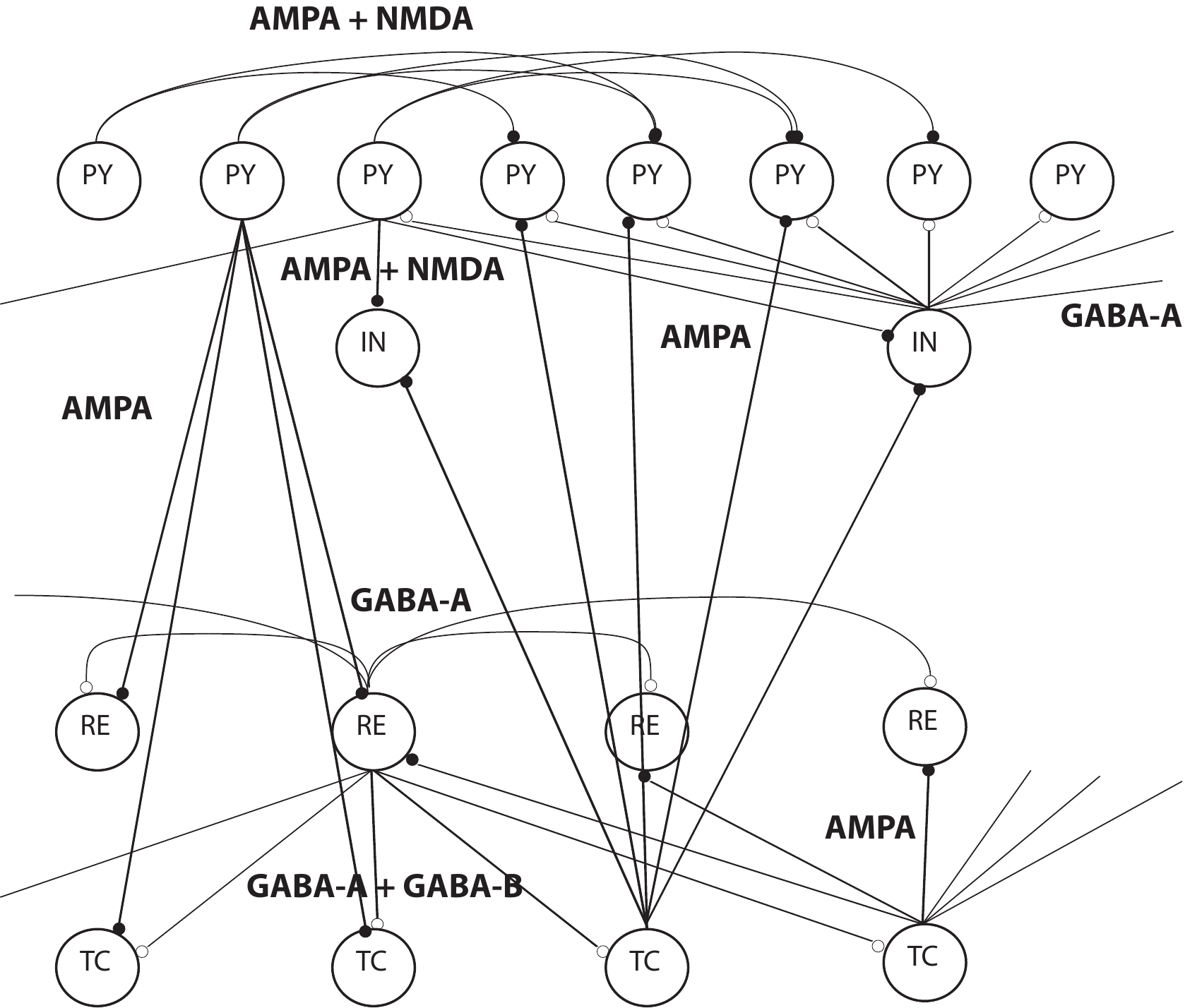}
\end{center}
\caption{Network model for the thalamocortical interactions, which included four layers of neurons. Thalamocortical (TC), reticular nucleus (RE) of the thalamus, cortical pyramidal (PY) and inhibitory (IN) neurons. Small solid dots indicate excitatory synapses. Small open dots indicate inhibitory synapses.}
\label{thalamoModel}
\end{figure}
To test and benchmark the differential covariance-based methods in a more realistic model, we simulated neural signals from a Hodgkin-Huxley based spiking neural network.
The thalamocortical model used in this study was based on several
previous studies, which were used to model spindle and slow wave activity \citep{bazhenov02-8691, chen12-3987, bonjean11-9124}.
A schematic of the thalamocortical model in this work is shown in Fig.~\ref{thalamoModel}.
As shown, the thalamocortical model was structured as a one-dimensional, multi-layer array of cells. 
The thalamocortical network consisted of 50 cortical pyramidal
(PY) neurons, 10 cortical inhibitory (IN) neurons, 10 thalamic relay (TC)
neurons and 10 reticular (RE) neurons.
The connections between the 50 PY neurons follow the pattern in the passive neuron model and are shown in Fig.~\ref{HHold}A. For the rest of the connection types, a neuron connects to all target neurons within the radius listed in Table~\ref{tab_1} \citep{bazhenov02-8691}. The network is driven by spontaneous oscillations. Details of the model are explained in appendix~\ref{thalamo}.

\begin{table}[h]
\caption{Connectivity properties
}
\label{tab_1}
\resizebox{\columnwidth}{!}{
\begin{tabular}{c||c|c|c|c|c|c|c|c|c|c}
    & PY$\rightarrow$TC & PY$\rightarrow$RE & TC$\rightarrow$PY & TC$\rightarrow$IN    & PY$\rightarrow$IN & IN$\rightarrow$PY & RE$\rightarrow$RE & TC$\rightarrow$RE & RE$\rightarrow$TC \\ \hline  \hline
  Radius & 8 & 6 & 15 & 3  & 1 & 5 & 5 & 8 & 8\\ \hline
  %SC & CA & CA & CA & CA  & CA, CN &  CG$_A$ & CG$_A$ & CA & CG$_A$, CG$_B$ \\ \hline
%\multicolumn{11}{l}{
%\begin{minipage}[t]{1.4\textwidth}
%Top rows (FO: fanout rate): Fanouts used for connections in the thalamocortical network model. Multiple numbers within a single cell represent parameter changes with increased fanout. Bottom row (SC: synaptic current): synaptic currents of network connectivity. 
%\end{minipage}
%}
  %synaptic current & AMPA & AMPA & AMPA & AMPA & AMPA, NMDA & AMPA, NMDA & GABA$_A$ & GABA$_A$ & AMPA & GABA$_A$, GABA$_B$ \\ \hline 
\end{tabular}
}
\end{table}

For each simulation, we simulated the network for 600 secs. The data were sampled at 1000 Hz.

\subsubsection{Local field potential (LFP) model}

\begin{figure}[hbtp]
\begin{center}
\includegraphics[width=\textwidth]{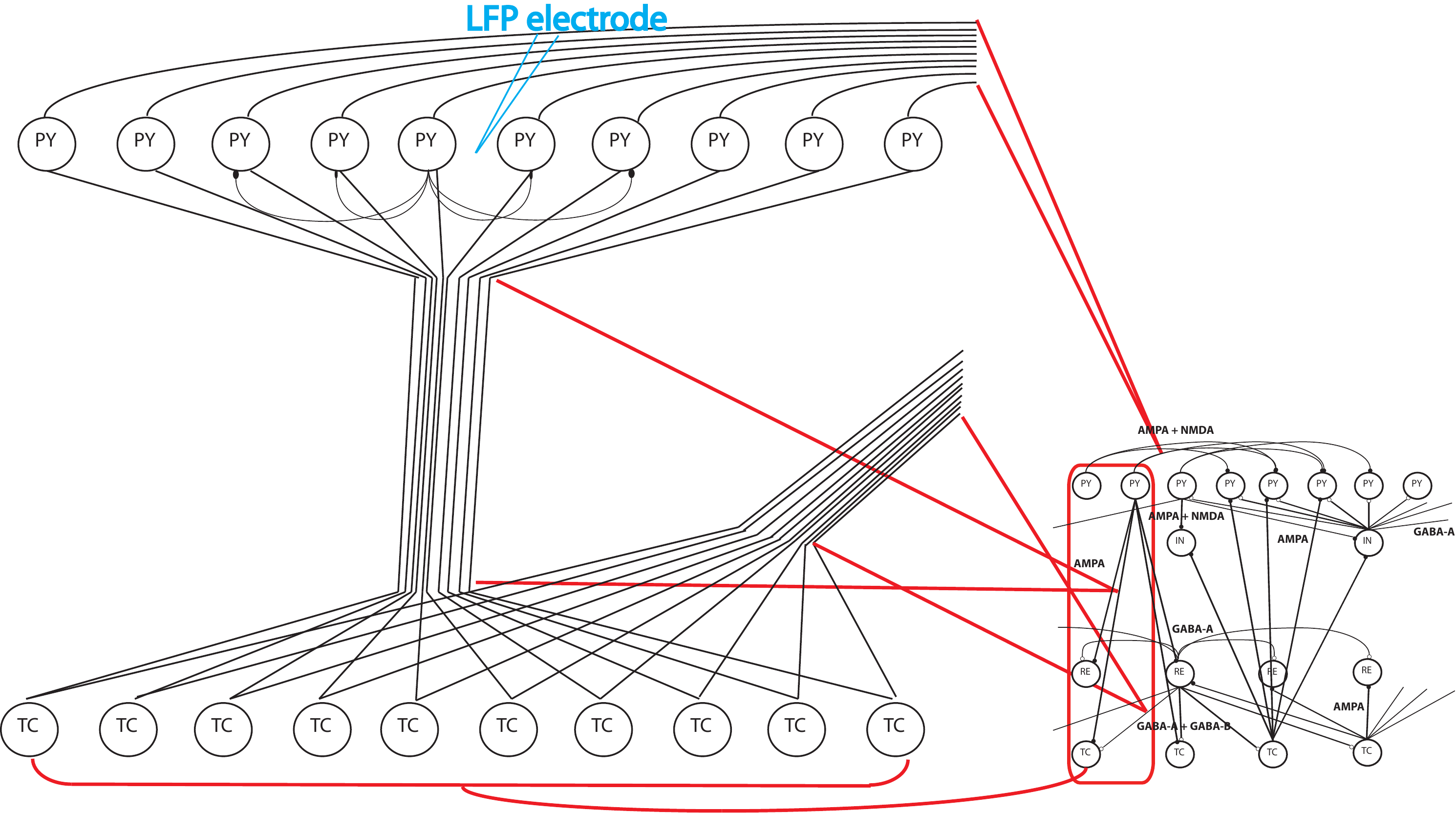}
\end{center}
\caption{The LFP model was transformed from the thalamocortical model. 
Shown in this figure, we used the TC $\rightarrow$ PY connection in the red box as an example. 
Each neuron in the original thalamocortical model was expanded into a sub-network of 10 neurons. 
Each connection in the original thalamocortical model was transformed into 10 parallel connections between two sub-networks.
Moreover, sub-networks transformed from PY neurons have local connections. 
Each PY neuron in this sub-network connects to its 2 nearest neighbors on each side.
We put a LFP electrode at the center of each PY sub-network.
The electrode received neurons' signals inversely proportional to the distance.}
\label{2layer}
\end{figure}

To simulate local field potential data, we expanded the previous thalamocortical model by 10 times (Fig.~\ref{2layer}). Each connection in the previous network (Fig.~\ref{thalamoModel}) is transformed into a fiber bundle connecting 10 pairs of neurons in a sub-network. For cortical pyramidal neurons (PY), we connect each neuron to its 4 neighboring neurons inside the sub-network. Other settings for the neurons and the synapses are the same as the previous thalamocortical model. We simulated the network for 600 secs. The data were sampled at 1000 Hz.

We plant a LFP electrode at the center of each of the 50 PY neuron local circuits. 
The distance between the 10 local neurons is $\SI{100}{\micro\meter}$, and the distance between the PY sub-networks is $\SI{1}{\centi\meter}$. The LFP is calculated according to the ``standard model'' as previously mentioned in \cite{bonjean11-9124, destexhe1998spike, nunez2005electric}. 
The LFPs are mainly contributed by elongated dendrites of the cortical pyramidal neurons. In our model, each cortical pyramidal neuron has a $\SI{2}{\milli\meter}$ long dendrite. 

For each LFP $S_i$,

\begin{equation}
S_i = \sum_j(\frac{I_{syn}}{r^i_d}-\frac{I_{syn}}{r^i_s})_j
\end{equation}

Where the sum is taken over all excitatory cortical neurons. $I_{syn}$ is the post-synaptic current of neuron $j$. $r_d$ is the distance from the electrode to the center of the dendrite of neuron $j$. $r_s$ is the distance from the electrode to the soma of neuron $j$.

\subsubsection{Calcium imaging model}
\cite{vogelstein2009spike} proposed a transfer function between spike trains and calcium florescence signals.
\begin{equation}
\begin{array}{c}
[Ca]_t - [Ca]_{t-1} = -\frac{\Delta t}{\tau_{Ca}}[Ca]_{t-1} + A_{Ca}n_t \\
F = \frac{[Ca]}{[Ca]+K_d} + \eta_t
\end{array}
\end{equation}

Where $A_{Ca}=\SI{50}{\micro\nauticalmile}$ is a step influx of calcium molecules at each action potential. $n_t$ is the number of spikes at each time step. $K_d = \SI{300}{\micro\nauticalmile}$ is the saturation concentration of calcium. $\eta_t$ is a Gaussian noise with a standard deviation of $0.000003$. 
Since our data is sampled at 1000 Hz, we can resolve every single action potential. So in our data, there is no multiple spikes at one time step. 
$\tau_{Ca}=\SI{1}{\second}$ is the decay constant for calcium molecules. To maintain the information in the differential signal, instead of setting a hard cutoff value and transform the intracellular voltages to binary spike trains, we use a sigmoid function to transform the voltages to the calcium influx activation parameter ($n_t$).

\begin{equation}
n_t = \frac{1}{1+e^{-(V(t)-V_{thre})}}
\end{equation}

Where $V_{thre} = \SI{-50}{\milli\volt}$ is the threshold potential.

In real experiments, we can only image a small percentage of neurons in the brain. Therefore, we simulated HH-based networks of 1000 neurons and only record from 50 neurons. We used the 4 connection patterns (normal-1 $\sim$ normal-4) provided on-line (\url{ https://www.kaggle.com/c/connectomics} ) \citep{stetter2012model}. Briefly, the networks have 80\% excitatory neurons and 20\% inhibitory neurons. The connection probability is 0.01, i.e. one neuron connects to about 10 neurons, so it is a sparsely connected network. 

Similar to our thalamocortical model, we used AMPA and NMDA synapses for the excitatory synapses, and GABA synapses for the inhibitory synapses. The simulations ran 600secs and were sampled at 1000Hz. 
The intracellular voltages obtained from the simulations were then transferred to calcium florescence signals and down sampled to 50 Hz. For each of the 4 networks, we conducted 25 recordings. Each recording contains 50 randomly selected neurons' calcium signals.

For accurate estimations, the differential covariance-based methods require the reconstructed action potentials from the calcium imaging. 
While this is an active research area and many methods have been proposed \citep{quan2010method, rahmati2016inferring}, In this study, we simply reversed the transfer function.
By assuming the transfer function from action potentials to calcium fluorescence signals is known, we can reconstruct the action potentials as:

Given $\hat{F}$ as the observed fluorescence signal
\begin{equation}
\begin{array}{c}
\hat{[Ca]}=\hat{F}*K_d/(1-\hat{F}) \\
\hat{n_t}=(\hat{d[Ca]}+1/\tau_{Ca}[Ca]_t)/(A/\Delta t) \\
\hat{V}=log(1/n_t-1)
\end{array}
\end{equation}

\subsection{Differential covariance-based Methods}
In this section, we introduce a new class of methods to estimate the functional connectivity of neurons (code is provided on-line at \url{https://github.com/tigerwlin/diffCov}).

\subsubsection{Step 1: differential covariance}
\label{diffCov_algo}
The input to the method, $V(t)$, is a $N \times T$ neural recording dataset. N is the number of neurons/channels recorded, and T is the number of data samples during recordings. We compute the derivative of each time series with $dV(t)=(V(t+1)-V(t-1))/(2dt)$. Then, the covariance between V(t) and dV(t) is computed and denoted as $\Delta_C$, which is a $N \times N$ matrix defined as the following:
\begin{equation}
\Delta_{C_{i,j}}=cov(dV_i(t),V_j(t))
\end{equation}
where $dV_i(t)$ is the differential signal of neuron/channel $i$, $V_j(t)$ is the signal of neuron/channel $j$, and $cov()$ is the sample covariance function for two time series.
In appendix~\ref{diffCovMath}, we provide a theoretical proof about why the differential covariance estimator generates less false connections than the covariance-based methods.

\subsubsection{Step 2: applying partial covariance method}
\label{partial_diffCov}

As previously mentioned in \cite{stevenson2008inferring}, one problem of the correlation method is the propagation of correlation. 
Here we designed a customized partial covariance algorithm to reduce this type of error in our methods.
We use $\Delta_{P_{i,j}}$ to denote the differential covariance estimation after applying the partial covariance method. 

Using the derivation from appendix~\ref{sec_PCOV}, we have:
\begin{equation}
\Delta_{P_{i,j}} = \Delta_{C_{i,j}} - COV_{j,Z}\cdot COV_{Z,Z}^{-1}\cdot \Delta_{C_{i,Z}}^T
\label{partDiffCov}
\end{equation}
Where $Z$ is a set of all neurons/channels except $\{i, j\}$.

$$Z=\{1, 2, ..., i-1, i+1,...,j-1,j+1,...N\}$$
 
$\Delta_{C_{i,j}}$ and $\Delta_{C_{i,Z}}$ were computed from section~\ref{diffCov_algo}, and $COV_{Z,Z}$ is the covariance matrix of set $Z$. $COV_{j,Z}$ is a flat matrix denoting the covariance of neuron $j$ and neurons in set $Z$. $\cdot$ is the matrix dot product.

As explained in appendix ~\ref{sec_PCOV}, the partial covariance of the differential covariance is not equivalent to the inverse of the differential covariance estimation. The two are only equivalent for the covariance matrix and when the partial correlation is controlling on all variables in the observed set. In our case, the differential covariance matrix is non-symmetric because it is the covariance between recorded signals and their differential signals. We have signals and differential signals in our observed set, however, we are only controlling on the original signals for the partial covariance algorithm. Due to these differences, we developed this customized partial covariance algorithm (Eq.~\ref{partDiffCov}), which performs well for neural signals in the form of Eq.~\ref{passiveModel}.

\subsubsection{Step 3: sparse latent regularization}

Finally, we applied the sparse latent regularization method to partial covariance version of the differential covariance \citep{chandrasekaran2011rank, yatsenko2015improved}. As explained in appendix~\ref{SLreg}, in the sparse latent regularization method, we made the assumption that there are observed neurons and unobserved common inputs in a network. 
If the connections between the observed neurons are sparse and the number of unobserved common inputs is small, this method can separate the covariance into two parts and the sparse matrix is the intrinsic connections between the observed neurons.

Here we define $\Delta_S$ as the sparse result from the method, and $L$ as the low-rank result from the method. 

Then by solving:
\begin{equation}
\argmin\limits_{\Delta_S,L} ||\Delta_S||_1 + \alpha*tr(L)
\end{equation}
under the constraint that 
\begin{equation}
\Delta_P =\Delta_S+L
\end{equation}

Where, $||\ ||_1$ is the L1-norm of a matrix, and $tr()$ is the trace of a matrix. $\alpha$ is the penalty ratio between the L1-norm of $\Delta_S$ and the trace of L. It was set to $1/\sqrt{N}$ for all our estimations. $\Delta_P$ is the partial differential covariance computed from section~\ref{partial_diffCov}. 
We receive a sparse estimation, $\Delta_S$, of the connectivity.

\subsubsection{Computing derivative}

In differential covariance, computing the derivative using $dV(t)=(V(t+1)-V(t-1))/(2dt)$ provides better estimation results than using $dV(t+1)=(V(t+1)-V(t))/dt$. To elaborate this point, we first explain our method's connection to the autoregressive (AR) method. 

Detailed discussion of the multivariate autoregressive model and its estimators has been discussed in \cite{harrison2003multivariate}.
In discrete space, our differential covariance estimator is similar to the mean squared error (MSE) estimator of the AR model. 
Following the definition in Eq.~\ref{passiveModel}, we have:
\begin{equation}
\begin{array}{c}
C\frac{dV(t)}{dt} =G \cdot V(t) + \mathcal{N}
\end{array}
\end{equation}
where, $V(t)$ are neurons' membrane voltages. $G$ is the connection matrix that describes the conductance between each pair of neurons. $\mathcal{N}$ is the Gaussian noise.

For $dV(t)=(V(t+1)-V(t-1))/(2dt)$, we note here:
\begin{equation}
\begin{array}{c}
C\frac{dV(t)}{dt} = C\frac{V(t+1)-V(t-1)}{2\Delta t}=G \cdot V(t) + \mathcal{N}	\\
V(t+1)=\frac{2\Delta t}{C} G \cdot V(t)+V(t-1)+ \frac{2\Delta t}{C} \mathcal{N}
\end{array}
\end{equation}
As explained in \cite{harrison2003multivariate}, in AR model, the MSE estimator of $G$ is $\frac{(V(t+1)-V(t-1))V(t)^T}{V(t)V(t)^T}$. We note that the numerator of this MSE estimator is our differential covariance estimator. Therefore, in this case, the model we proposed is equivalent to a regularized AR(2) model, where the transition matrix of the $2^{nd}$ order is restricted to be an identity matrix.

In the $dV(t)=(V(t+1)-V(t))/dt$ case, we note that the differential covariance estimator is similar to an AR(1) estimator:

\begin{equation}
\begin{array}{c}
C\frac{dV(t)}{dt} = C\frac{V(t+1)-V(t)}{\Delta t}=G \cdot V(t) + \mathcal{N}	\\
V(t+1)=(\frac{\Delta t}{C} G+I) \cdot V(t)+ \frac{\Delta t}{C} \mathcal{N}
\end{array}
\end{equation}

And the MSE estimator of $(\frac{\Delta t}{C} G+I)$ is $\frac{V(t+1)V(t)^T}{V(t)V(t)^T}$. Therefore:

\begin{align}
&\begin{aligned}
\frac{\Delta t}{C} G &= \frac{V(t+1)V(t)^T}{V(t)V(t)^T} - I \\
&= \frac{(V(t+1)-V(t))V(t)^T}{V(t)V(t)^T}\\
 \end{aligned}
\end{align}
where, the numerator of this MSE estimator is our differential covariance estimator.

As explained above, using different methods to compute the derivative will produce different differential covariance estimators and they are equivalent to estimators from different AR models for the connection matrix. In section~\ref{sec_results}, we show that the performance of these two estimators are significantly different.

\subsection{Performance quantification}

The performance of each method is judged by 4 quantified values. The first 3 values indicate the method's abilities to reduce the 3 types of false connections (Fig.~\ref{falseConn}). The last one indicates the method's ability to correctly estimate the true positive connections against all possible interference.

Let's define $G$ as the ground truth connectivity matrix, where:
\begin{equation}
    G_{i,j}= 
\begin{cases}
    1,  & \text{if neuron i projects to neuron j with excitatory synapse}\\
    -1,  &  \text{if neuron i projects to neuron j with inhibitory synapse}\\
    0,                 & \text{otherwise}
\end{cases}
\end{equation}

Then, we can use a 3-dimensional tensor to represent the false connections caused by common inputs. For example, neuron $j$ and neuron $k$ receive common input from neuron $i$:
\begin{equation}
    M_{i,j,k}= 
\begin{cases}
    1,  & \text{iff } G_{i,j}=1 \text{ and } G_{i,k}=1\\
    0,                 & \text{otherwise}
\end{cases}
\end{equation}

Therefore, we can compute a mask that labels all the type 1 false connections:
 
\begin{equation}
    %mask^1_{j,k} = \sum_{i \in \{\text{observed neurons}\}}M_{i,j,k}
    mask_{1_{j,k}} = \sum_{i \in \{\text{observed neurons}\}}M_{i,j,k}
\end{equation}

For the type 2 false connections (e.g. neuron $i$ projects to neuron $k$, then neuron $k$ projects to neuron $j$), the mask is defined as:
\begin{equation}
mask_{2_{i,j}} = \sum_{k \in \{\text{observed neurons}\}}G_{i,k}G_{k,j}
\end{equation}
or, in simple matrix notation:

\begin{equation}
mask_2 = G \cdot G
\end{equation}

Similar to $mask_1$, the false connections caused by unobserved common inputs is:
\begin{equation}
    mask_{3_{j,k}} = \sum_{i \in \{\text{unobserved neurons}\}}M_{i,j,k}
\end{equation}

Lastly, $mask_4$ is defined as:

\begin{equation}
    mask_{4_{i,j}} = 
      \begin{cases}
        1,  & \text{if } G_{i,j}=0\\
        0,                 & \text{otherwise}
\end{cases}
\end{equation}

Given a connectivity matrix estimation result: $Est$,
the 4 values for the performance quantification are computed as the area under the ROC curve for two sets:
the true positive set and the false positive set.
\begin{equation}
\begin{array}{c}
|Est_{i,j}| \in \{\text{true positive set}\}_k, \text{if } G_{i,j} \neq 0 \text{ and } mask_{k_{i,j}}=0 \\
|Est_{i,j}| \in \{\text{false positive set}\}_k, \text{if } mask_{k_{i,j}} \neq 0 \text{ and } G_{i,j}=0
\end{array}
\end{equation}
\section{Results}
\label{sec_results}
\begin{figure}[hbtp]
\begin{center}
\includegraphics[width=.8\textwidth]{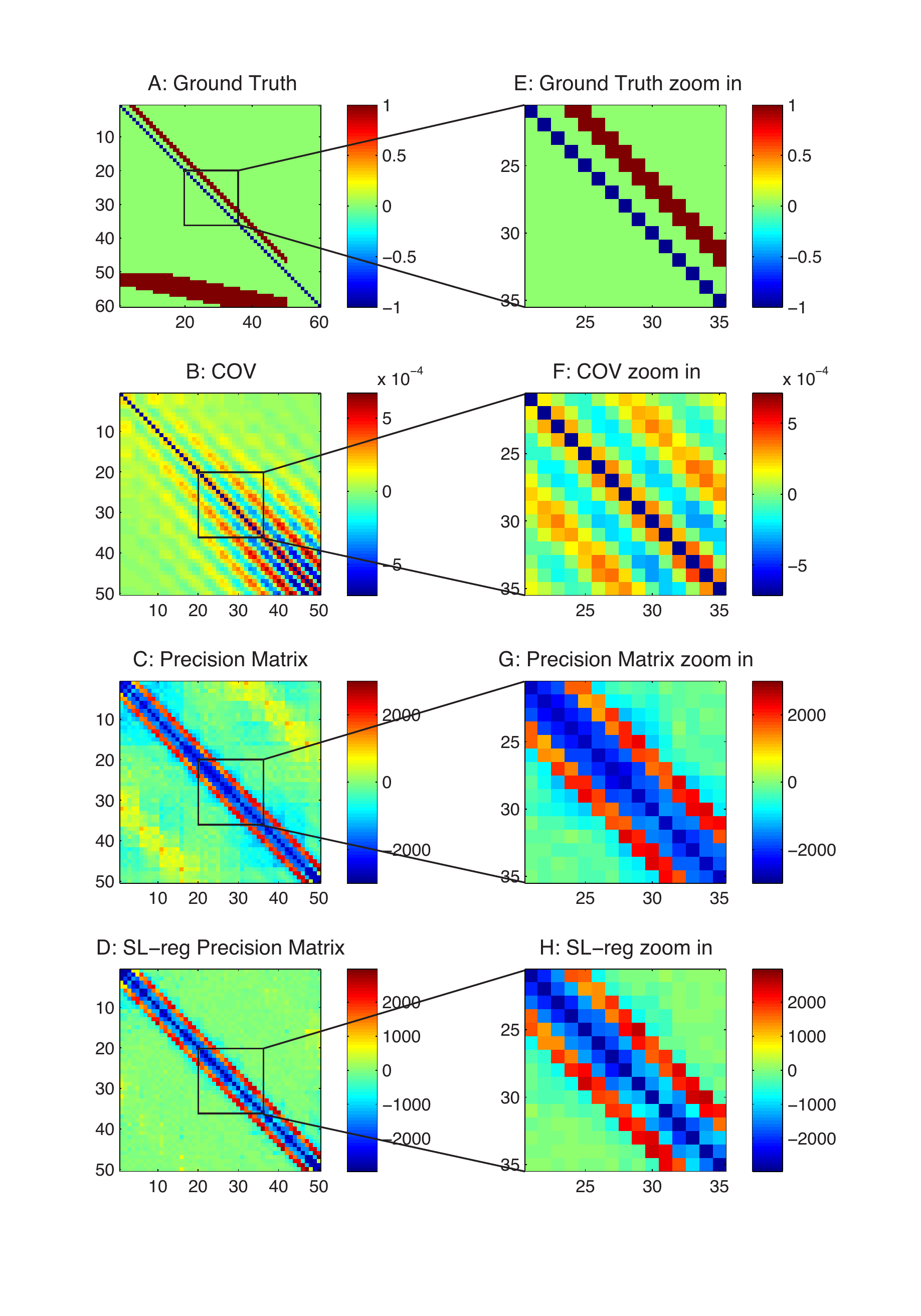}
\end{center}
\caption{Ground truth connections from a passive neuron model and estimations from correlation-based methods. A) Ground truth connection matrix. neurons 1-50 are observed neurons. neurons 51-60 are unobserved neurons that introduce common inputs. B) Estimation from the correlation method. C) Estimation from the precision matrix. D) Estimation from the sparse+latent regularized precision matrix. E) Zoom in of panel A. F) Zoom in of panel B. G) Zoom in of panel C. H) Zoom in of panel D.}
\label{circuitOld}
\end{figure}
\subsection{False connections in correlation-based methods}

When applied to neural circuits, the commonly used correlation-based methods produce systematic false connections. 
As shown, Fig.~\ref{circuitOld}~A is the ground truth of the connections in our passive neuron model (Neurons 1-50 are the observed neurons). Fig.~\ref{circuitOld}B is from the correlation method, Fig.~\ref{circuitOld}C is the precision matrix, and Fig.~\ref{circuitOld}D is the sparse+latent regularized precision matrix. As shown, all of these methods produce extra false connections. 

For the convenience of explanation, we define the diagonal strip of connections in the ground truth (first 50 neurons in  Fig.~\ref{circuitOld}A) as the -3 and -4 diagonal lines, because they are 3 and 4 steps away from the diagonal line of the matrix. As shown in Fig.~\ref{circuitOld}, all these methods produce false connections on the $\pm1$ diagonal lines. The precision matrix method (Fig.~\ref{circuitOld}C) also has square box shape false connections in the background. 

\begin{figure}[hbtp]
\begin{center}
\includegraphics[width=\textwidth]{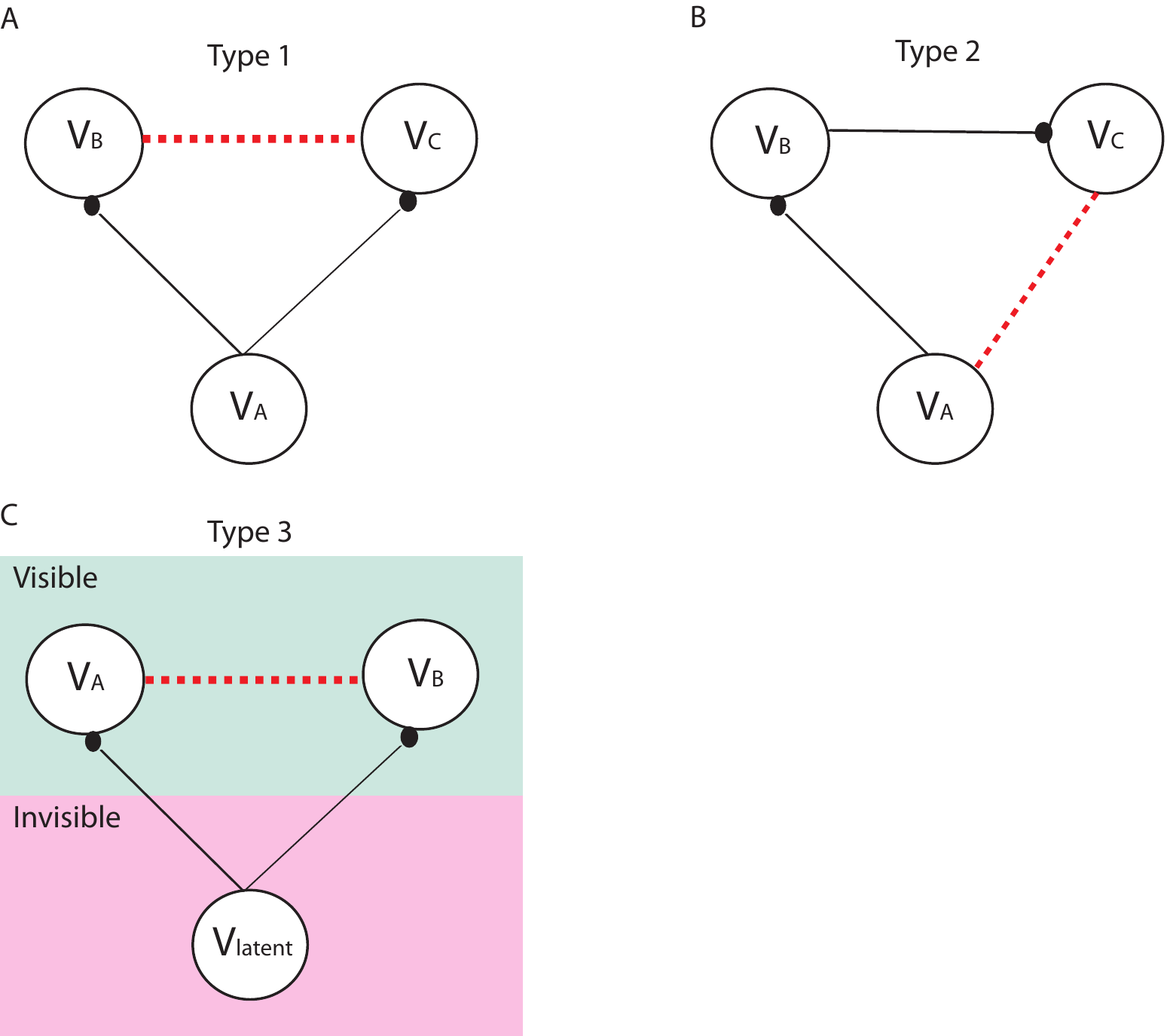}
\end{center}
\caption{Illustrations of the 3 types of false connections in the correlation-based methods. Solid lines indicate the physical wiring between neurons, and the black circles at the end indicate the synaptic contacts (i.e. the direction of the connections). The dotted lines are the false connections introduced by the correlation-based methods. A) Type 1 false connections, which are due to two neurons receiving the same synaptic inputs. B) Type 2 false connections, which are due to the propagation of correlation.  C) Type 3 false connections, which are due to unobserved common inputs.}
\label{falseConn}
\end{figure}

\subsubsection{Type 1 false connections}
\label{type1}

Shown in Fig.~\ref{falseConn}A, the type 1 false connections are produced because two neurons receive the same input from another neuron. The same synaptic current that passes into the two neurons generates positive correlation between the two neurons. However, there is no physiological connection between these two neurons. In our connection pattern (Fig.~\ref{circuitOld} A), we notice that two neurons next to each other receive common synaptic inputs, therefore there are false connections on the  $\pm1$ diagonal lines of the correlation-based estimations.

\subsubsection{Type 2 false connections}

Shown in Fig.~\ref{falseConn}B, the type 2 false connections are due to the propagation of correlation. Because one neuron $V_A$ connects to another neuron $V_B$ and neuron $V_B$ connects to another neuron $V_C$, the correlation method presents correlation between $V_A$ and $V_C$, which do not have a physical connection. This phenomenon is shown in Fig.~\ref{circuitOld}B as the extra diagonal strips. Shown in Fig.~\ref{circuitOld}C, this problem can be greatly reduced by the precision matrix/partial covariance method. 

\subsubsection{Type 3 false connections}

Shown in Fig.~\ref{falseConn}C, the type 3 false connections are also due to the common currents pass into two neurons. However, in this case, they are from the unobserved neurons. For this particular passive neuron model, it is due to the inputs from the 10 unobserved neurons (Neurons 51-60) as shown on Fig.~\ref{circuitOld}A. Because the latent neurons have broad connections to the observed neurons, they introduce a square box shape correlation pattern into the estimations (Fig.~\ref{circuitOld}C. Fig.~\ref{circuitOld}B also contains this error, but it is hard to see). Since, the latent neurons are not observable, partial covariance method cannot be used to regress out this type of correlation. On the other hand, the sparse latent regularization can be applied if the sparse and low-rank assumption is valid, and the sparse+latent regularized result is shown in Fig.~\ref{circuitOld}D. 
However, even after using this regularization, the correlation-based methods still leave false connections in Fig.~\ref{circuitOld}D.

%\FloatBarrier
\subsection{Estimations from differential covariance-based methods}
\begin{figure}[hbtp]
\begin{center}
\includegraphics[width=.7\textwidth]{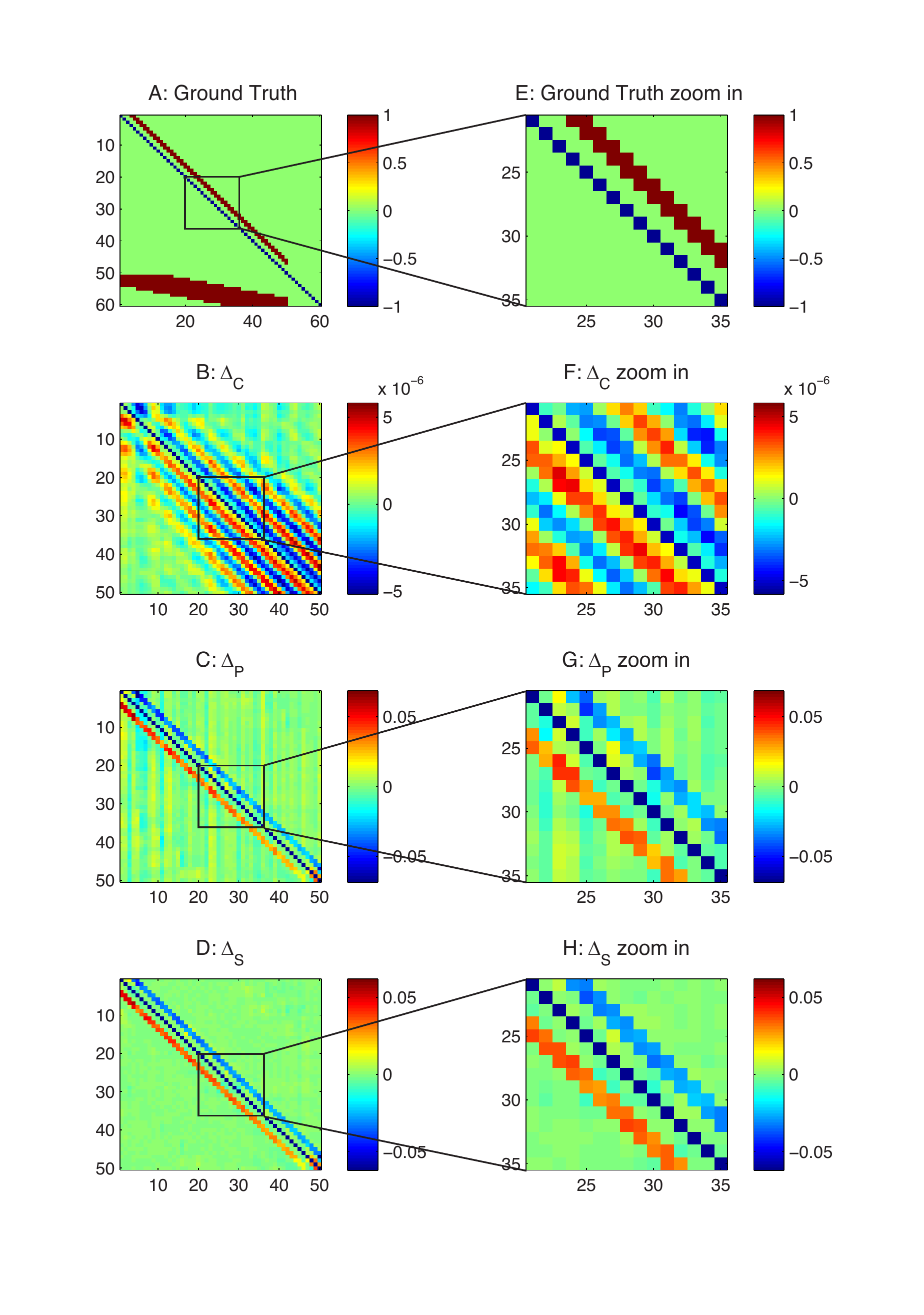}
\end{center}
\caption{differential covariance analysis of the passive neuron model. The color in B,C,D,F,G,H indicates direction of the connections. For element $A_{ij}$, warm color indicates $i$ is the sink, $j$ is the source, i.e. $i \leftarrow j$, and cool color indicates $j$ is the sink, $i$ is the source,  i.e. $i \rightarrow j$.  A) Ground truth connection matrix. B) Estimation from the differential covariance method. C) Estimation from the partial differential covariance method. D) Estimation from the sparse+latent regularized  partial differential covariance method. E) Zoom in of panel A. F) Zoom in of panel B. G) Zoom in of panel C. H) Zoom in of panel D.}
\label{circuitNew}
\end{figure}

\begin{figure}[hbtp]
\begin{center}
\includegraphics[width=\textwidth]{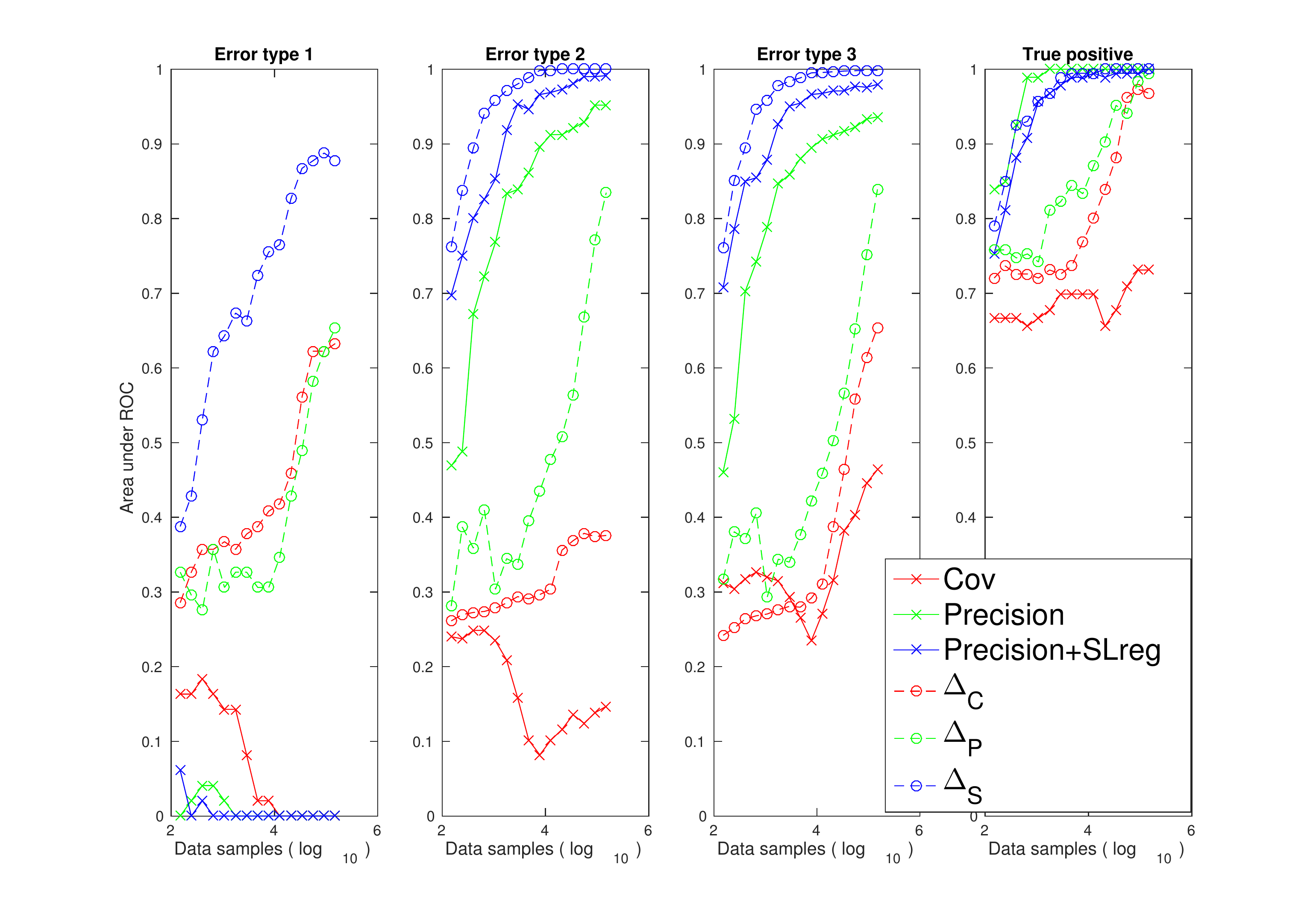}
\end{center}
\caption{Performance quantification (area under the ROC curve) of different methods with respect to their abilities to reduce the 3 types of false connections and their abilities to estimate the true positive connections using the passive neuron dataset.}
\label{circuitQuant}
\end{figure}

\begin{table}[h!]
\caption{Performance quantification (area under the ROC curve) of different methods with respect to their abilities to reduce the 3 types of false connections and their abilities to estimate the true positive connections under 5 different passive neuron model settings.
}
\label{circuitTable}
\resizebox{.75\columnwidth}{!}{
%\resizebox{!}{.7\textwidth}{
\begin{tabular}{c||c|c|c|c|c|c}
    & Cov & Precision & Precision+SL-reg & $\Delta_C$    & $\Delta_P$ & $\Delta_S$  \\ \hline  \hline
 cxcx34 g5  & &   &  &   &  &  \\ \hline
  Error 1 & 0 & 0 & 0 & 0.6327  & 0.3469 & 0.8776 \\ \hline
  Error 2 &   0.1469  &  0.9520  &  0.9915  &  0.3757  &  0.8347  &  1.0000 \\ \hline
  Error 3 & 0.4638  &  0.9362  &  0.9797  &  0.6541 &   0.8391  &  0.9986 \\ \hline
  True Positive & 0.7312 & 1.0000 & 1.0000 &  0.9677 & 0.9946 &  1.0000\\ \hline
 cxcx34 g30 &  &  &  &   &  &  \\ \hline
Error 1 &0    &     0    &     0 &   0.0510  &  0.5816  &  0.9490	\\ \hline
Error 2 &0.0056  &  0.8927  &  0.9972  &  0.2881  &  0.9548  &  1.0000	\\ \hline
Error 3 &0.2164  &  0.9188  &  0.9942  &  0.5430  &  0.9662  &  1.0000\\ \hline
True Positive &    0.5591 &   1.0000  &  0.9892  &  0.6559  &  1.0000  &  1.0000	\\ \hline
 cxcx34 g50 &  &  &  &   &  &  \\ \hline
  Error 1 & 0     &    0    &     0    &     0   & 0.2041  &  0.6531 \\ \hline
  Error 2 &  0  &  0.7034  &  0.9944  &  0.0523   & 0.9054   & 1.0000 \\ \hline
  Error 3 & 0.3179  &   0.8000  &  0.9894  &  0.4145  &  0.9309  &  1.0000 \\ \hline
  True Positive & 0.9140  &  1.0000  &  0.9946  &  0.9516  &  0.9785  &  1.0000 \\ \hline
 cxcx56789 g5 &  &  &  &   &  &  \\ \hline
  Error 1 & 0  &  0.0053  &  0.0053   & 0.6895   & 0.6263 &   0.8526 \\ \hline
  Error 2 & 0.1896   & 0.6229  &  0.7896  &  0.5240  &  0.7896  &  0.9938 \\ \hline
  Error 3 &  0.3573  &  0.6085  &  0.7659  &  0.6957 &   0.7591 &   0.9817 \\ \hline
  True Positive & 0.6884  &  0.9442  &  0.6674   & 0.9930  &  0.9605  &  0.9837\\ \hline
 cxcx56789 g50 &  &  &  &   &  &  \\ \hline
  Error 1 &  0     &    0    &     0  &  0.0263   & 0.2816   & 0.6842 \\ \hline
  Error 2 & 0.1083  &  0.5312  &  0.8240  &  0.2844   & 0.6990  &  0.9979\\ \hline
  Error 3 &  0.4256  &  0.4927  &  0.7762  &  0.5091  &  0.7116  &  0.9835\\ \hline
  True Positive & 0.9256  &  0.9116  &  0.6698  &  0.9395  &  0.9279  &  0.9419 \\ \hline
\end{tabular}
}
\end{table}

Comparing the ground truth connections in Fig.~\ref{circuitNew}A with our final estimation in Fig.~\ref{circuitNew}D, we see that our methods essentially transformed the connections in the ground truth into a map of sources and sinks in a network. An excitatory connection, $i \rightarrow j$, in our estimations have negative value for $\Delta_{S_{ij}}$ and positive value for $\Delta_{S_{ji}}$, which means the current is coming out of the source $i$, and goes into the sink $j$. We note that there is another ambiguous case, an inhibitory connection $j \rightarrow i$, which produces the same results in our estimations. Our methods can not differentiate these two cases, instead, they indicate sources and sinks in a network.

\subsubsection{The differential covariance method reduces type 1 false connections}

By comparing Fig.~\ref{circuitOld} B with Fig.~\ref{circuitNew} B, we see that the type 1 false connections on the $\pm 1$ diagonal lines of Fig.~\ref{circuitOld} B is reduced in Fig.~\ref{circuitNew} B. 
This is reflecting the theorems we proved in appendix~\ref{diffCovMath}, in particular theorem 5, which shows that the strength of the type 1 false connections is reduced in the differential covariance method by a factor of $g_l/g_{syn}$. 
Moreover, the differential covariance method's performance on reducing the type 1 false connections is quantified in  Fig.~\ref{circuitQuant}.

\subsubsection{The partial covariance method reduces type 2 false connections}

Second, we see that, due to  the propagation of correlation, there are extra diagonal strips in Fig.~\ref{circuitNew}B. 
These are removed in Fig.~\ref{circuitNew}C by applying the partial covariance method. And each estimator's performance for reducing type 2 false connections is quantified in Fig.~\ref{circuitQuant}.

\subsubsection{The sparse+latent regularization reduces type 3 false connections}
Third, 
 the sparse+latent regularization to remove the correlation introduced by the latent inputs. As mentioned in the method section, when the observed neurons' connections are sparse and the number of unobserved common inputs is small, the covariance introduced by the unobserved common inputs can be removed. As shown in Fig.~\ref{circuitNew}D, the external covariance in the background of Fig.~\ref{circuitNew}C is removed, while the true diagonal connections and the directionality of the connections are maintained. This regularization is also effective for correlation-based methods, but type 1 false connections maintain in the estimation even after applying this regularization (Fig.~\ref{circuitOld}D). Each estimator's performance for reducing type 3 false connections is quantified in Fig.~\ref{circuitQuant}.

\subsubsection{The differential covariance-based methods provide directionality information of the connections}
Using this passive neuron model, in appendix~\ref{directionality}, we provide a mathematical explanation for why the differential covariance-based methods provide directional information for the connections. Given an excitatory connection $g_{i \rightarrow j}$ (neuron i projects to neuron j), from Theorem 6 in appendix~\ref{directionality}, we  have:

\begin{equation}
\begin{array}{lc}
\Delta_{C_{j,i}} > 0\\
\Delta_{C_{i,j}} < 0\\
\end{array}
\end{equation}

However, we wish to note here that, there is another ambiguous setting that provides the same result, which is an inhibitory connection $g_{j \rightarrow i}$. Conceptually, the differential covariance indicates the current sources and sinks in a neural circuit, but the exact type of synapse is unknown.

\subsubsection{Performance quantification for the passive neuron dataset}

In Figure~\ref{circuitQuant}, we quantified the performance of the estimators for one example dataset. We see that, with the increase of the sample size, our differential covariance-based methods reduce the 3 types of false connections, while maintaining high true positive rates. Also as we apply more advanced techniques ($\Delta_C \rightarrow \Delta_P \rightarrow \Delta_S$), the estimator's performance increases in all 4 panels of the quantification indices. 
Although the precision matrix and the sparse+latent regularization help the correlation method reduce the type 2 and type 3 error, all correlation-based methods handle poorly of the type 1 false connections. 
We also note that the masks we used to quantify each type of false connections are not mutually exclusive (i.e. there are false connections that belong to more than one type of false connections). Therefore, in Figure~\ref{circuitQuant}, it seems like a regularization is reducing multiple types of false connections. For example, the sparse+latent regularization is reducing both type 2 and type 3 false connections.

In Table~\ref{circuitTable}, we provide quantified results (area under the ROC curve) for two different connection patterns (cxcx34, and cxcx56789) and three different conductance settings (g5, g30, and g50). We see that, the key results in Fig.~\ref{circuitQuant} are also generalized here. By applying more advanced techniques to the original differential covariance estimator ($\Delta_C \rightarrow \Delta_P \rightarrow \Delta_S$, the performance increases with respect to the 3 types of error, while the true positive rate is not sacrificed. We also note that, although the precision matrix and the sparse+latent regularization help the correlation method reduce the type 2 and the type 3 error, all correlation-based methods handle poorly of the type 1 error.

%\FloatBarrier
\subsection{Thalamocortical model results}

We further tested the methods in a more realistic Hodgkin-Huxley based model. Because the synaptic conductances in the Hodgkin-Huxley model are no longer  constants but become nonlinear dynamic functions, which depend on the pre-synaptic voltages,  the derivations above can only be considered as a first-order approximation. 
\begin{figure}[hbtp]
\begin{center}
\includegraphics[width=.8\textwidth]{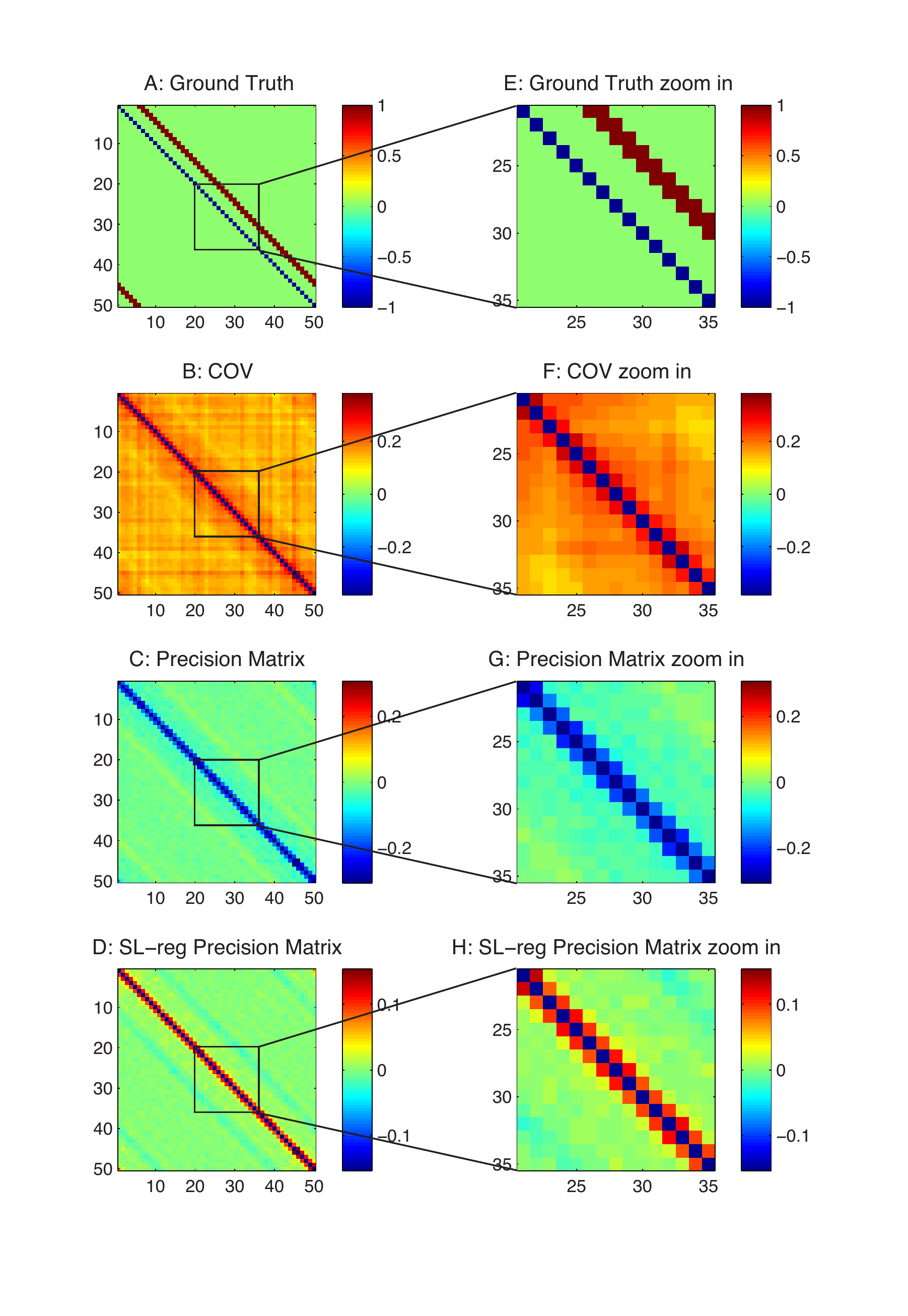}
\end{center}
\caption{Analysis of the thalamocortical model with correlation-based methods. A) Ground truth connections of the PY neurons in the thalamocortical model. B) Estimation from the correlation method. C) Estimation from the precision matrix method. D) Estimation from the sparse+latent regularized precision matrix method. E) Zoom in of panel A. F) Zoom in of panel B. G) Zoom in of panel C. H) Zoom in of panel D.}
\label{HHold}
\end{figure}

Shown in Fig.~\ref{HHold}A is the ground truth connections between the cortical neurons. These cortical neurons also receive latent inputs from and sending feedback currents to inhibitory neurons in the cortex (IN) and thalamic neurons (TC). For clarity of representation, these latent connections are not shown here, but the detailed connections are described in the Method section. 

Similar to the passive neuron model, in Fig.~\ref{HHold}B, the correlation method still suffers from those 3 types of false connections. 
As shown, the latent inputs generate false correlations in the background. 
And the $\pm1$ diagonal line false connections, which are due to the common currents, exist in all correlation-based methods (see  Fig.~\ref{HHold}B,C,D). Comparing Fig.~\ref{HHold}C, D, because the type 1 false connections are strong in the Hodgkin-Huxley based model, the sparse+latent regularization removed the true connections but kept these false connections in its final estimation.  
\begin{figure}[hbtp]
\begin{center}
\includegraphics[width=.7\textwidth]{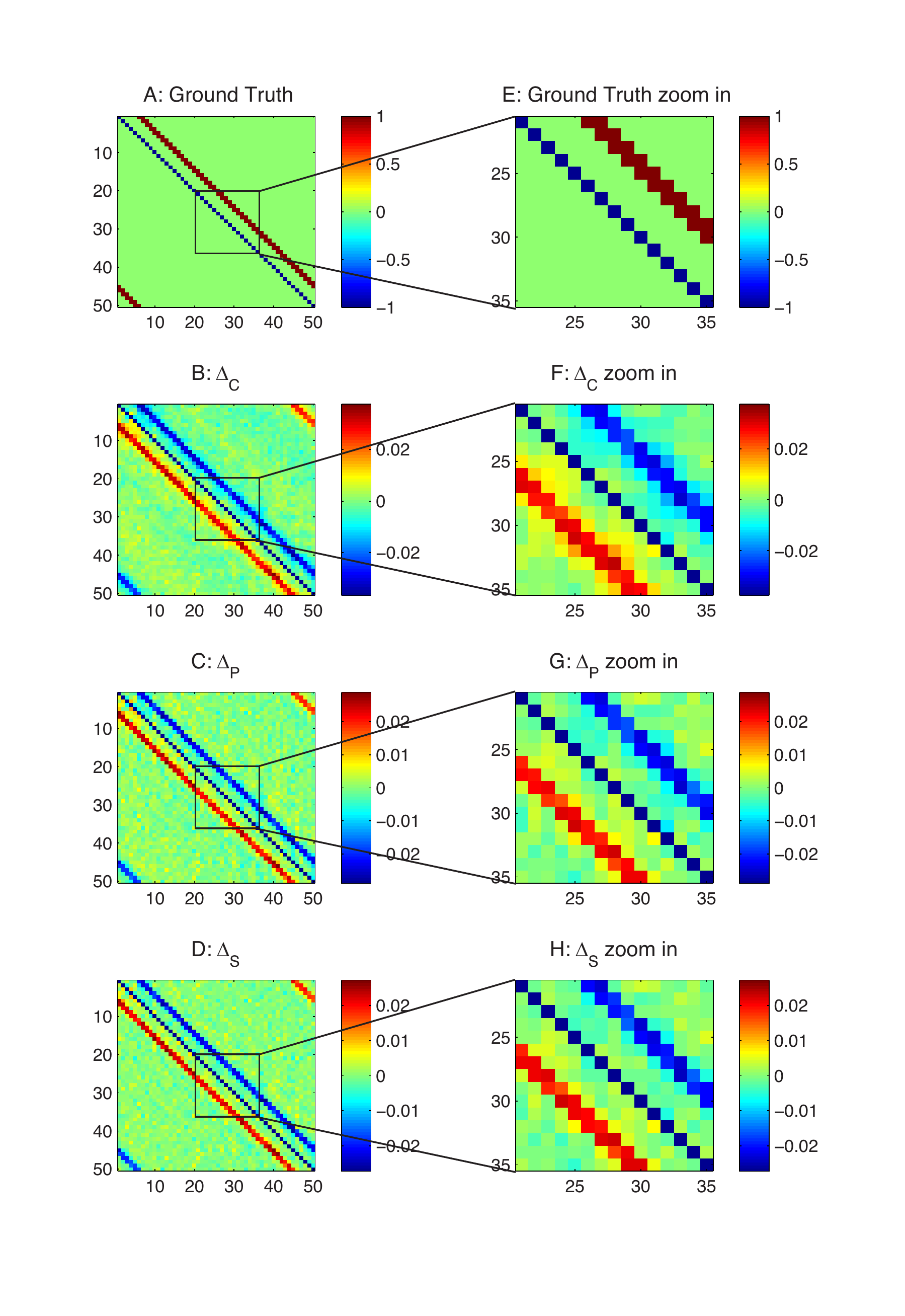}
\end{center}
\caption{Analysis of the thalamocortical model with differential covariance-based methods. The color in B,C,D,F,G,H indicates direction of the connections. For element $A_{ij}$, warm color indicates $i$ is the sink, $j$ is the source, i.e. $i \leftarrow j$, and cool color indicates $j$ is the sink, $i$ is the source,  i.e. $i \rightarrow j$. A) Ground truth connection matrix. B) Estimation from the differential covariance method. C) Estimation from the  partial differential  covariance method. D) Estimation from the sparse+latent regularized partial differential covariance method. E) Zoom in of panel A. F) Zoom in of panel B. G) Zoom in of panel C. H) Zoom in of panel D.}
\label{HHnew}
\end{figure}

\begin{figure}[hbtp]
\begin{center}
\includegraphics[width=\textwidth]{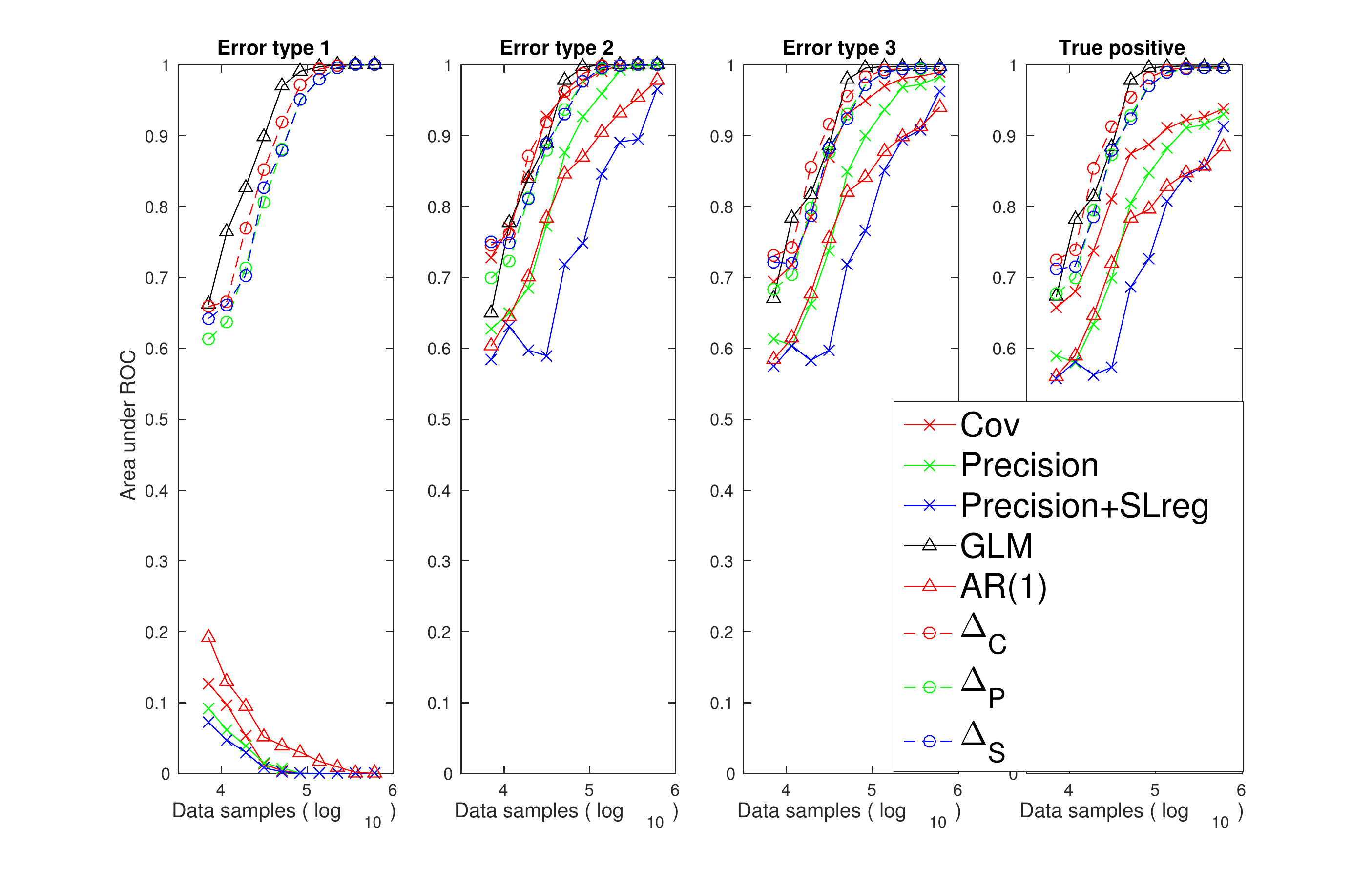}
\end{center}
\caption{Performance quantification (area under the ROC curve) of different methods with respect to their abilities to reduce the 3 types of false connections and their abilities to estimate the true positive connections using the thalamocortical dataset.}
\label{HHQuant}
\end{figure}

As shown in Fig.~\ref{HHnew}B, differential covariance method reduces the type 1 false connections. Then in Fig.~\ref{HHnew}C, the partial differential covariance method reduces type 2 false connections in Fig.~\ref{HHnew}B (yellow color connections around the red strip in Fig.~\ref{HHnew}B). Finally, in Fig.~\ref{HHnew}D, the sparse latent regularization removes the external covariance in the background of Fig.~\ref{HHnew}C. The current sources (positive value, red color) and current sinks (negative value, blue color) in the network are also indicated on our estimators. 

In Fig.~\ref{HHQuant}, each estimator's performance on each type of false connections is quantified. In this example, our differential covariance-based methods achieve similar performance to the GLM method.

\subsection{Simulated LFP results}

\begin{figure}[hbtp]
\begin{center}
\includegraphics[width=0.65\textwidth]{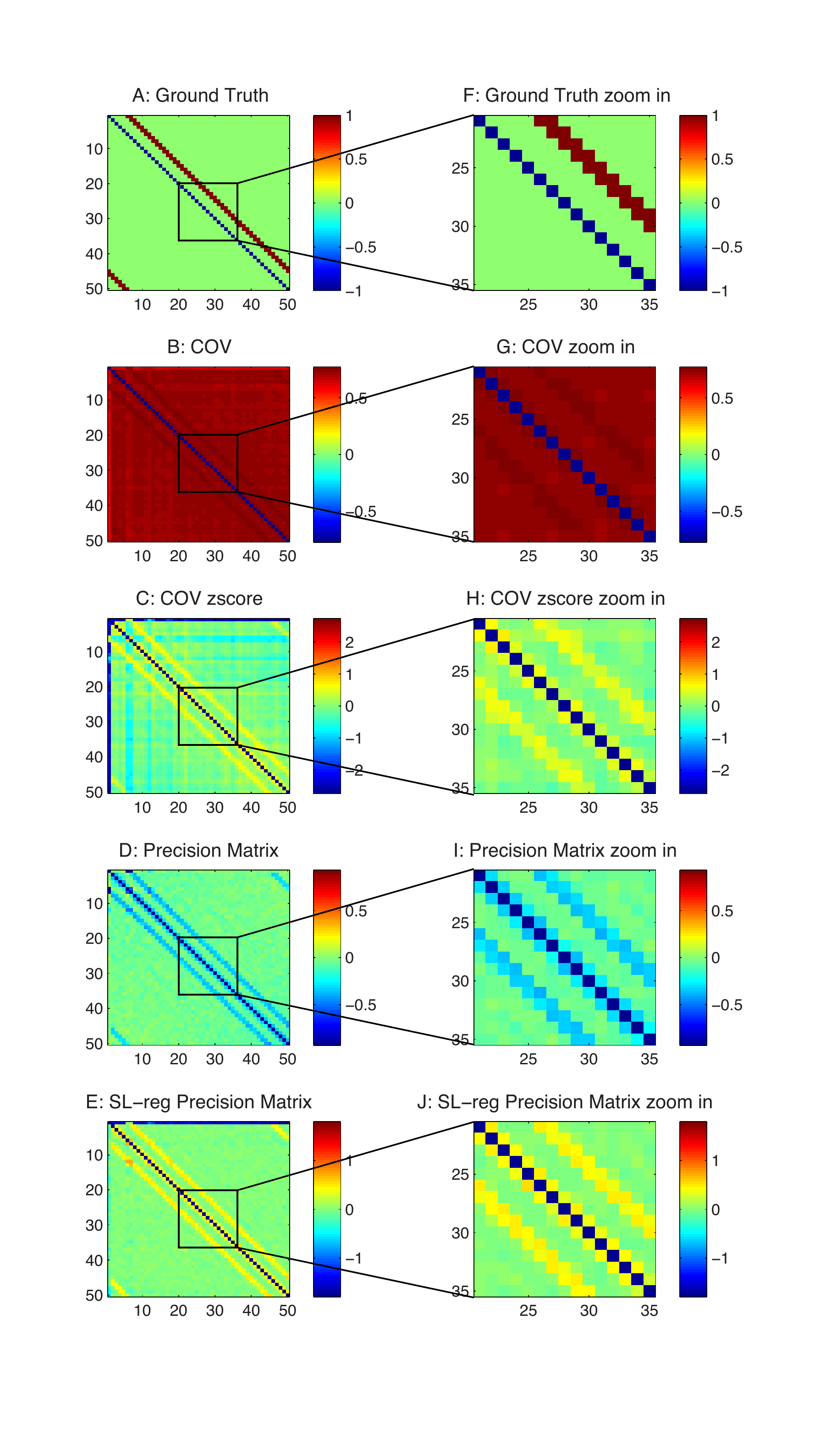}
\end{center}
\caption{Analysis of the simulated LFP data with correlation-based methods. A) Ground truth connection matrix B) Estimation from the correlation method. C) z-score of the correlation matrix. D) Estimation from the precision matrix method. E) Estimation from the sparse+latent regularized precision matrix method. F) Zoom in of panel A. G) Zoom in of panel B. H) Zoom in of panel C. I) Zoom in of panel D. J) Zoom in of panel E.}
\label{lfpOld}
\end{figure}

\begin{figure}[!htb]
\begin{center}
\includegraphics[width=.7\textwidth]{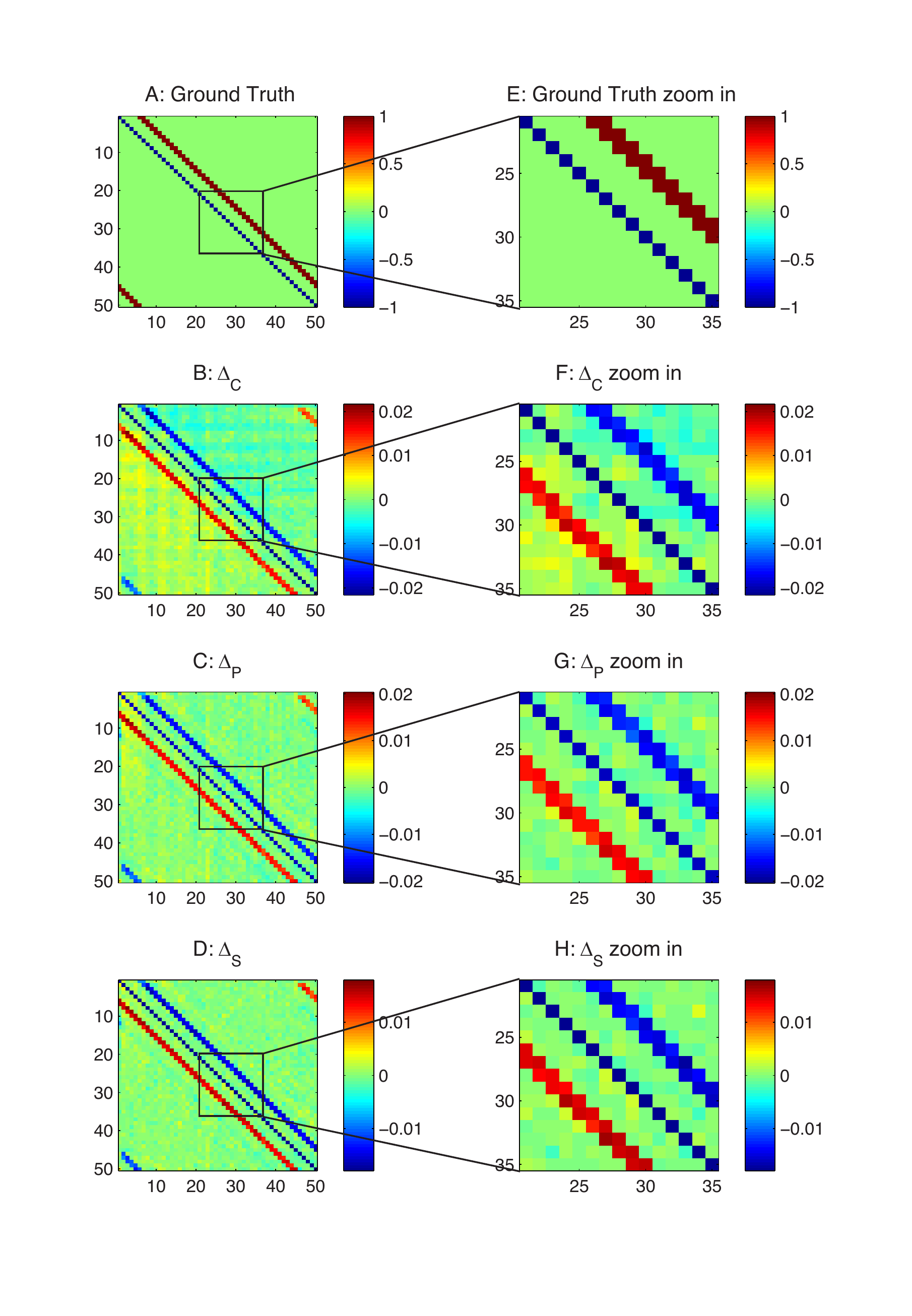}
\end{center}
\caption{Analysis of the simulated LFP data with differential covariance-based methods. The color in B,C,D,F,G,H indicates direction of the connections. For element $A_{ij}$, warm color indicates $i$ is the sink, $j$ is the source, i.e. $i \leftarrow j$, and cool color indicates $j$ is the sink, $i$ is the source,  i.e. $i \rightarrow j$. A) Ground truth connection matrix. B) Estimation from the differential covariance method. C) Estimation from the  partial differential covariance method. D) Estimation from the sparse+latent regularized partial differential covariance method. E) Zoom in of panel A. F) Zoom in of panel B. G) Zoom in of panel C. H) Zoom in of panel D.}
\label{lfpNew}
\end{figure}

\begin{figure}[!htb]
\begin{center}
\includegraphics[width=\textwidth]{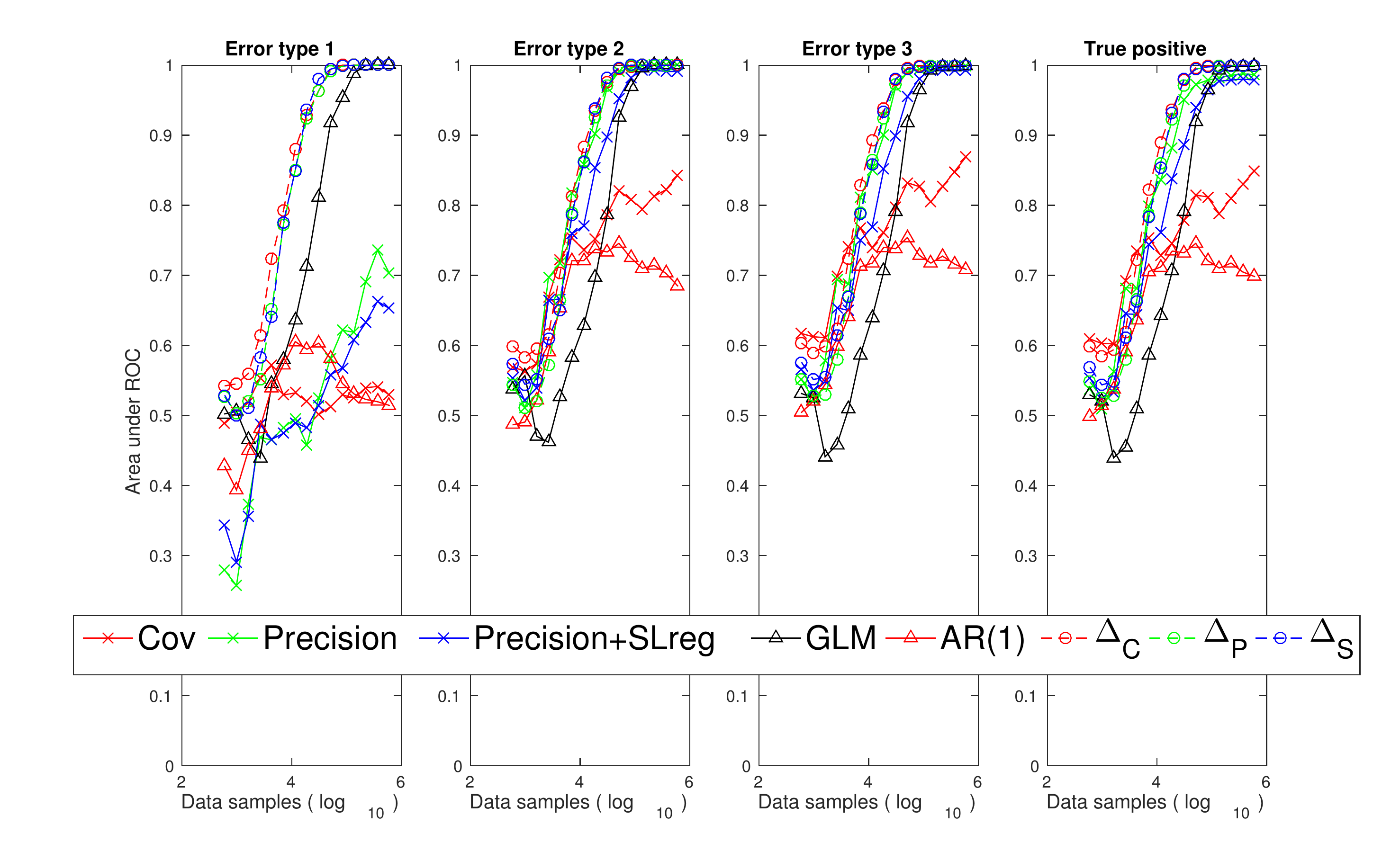}
\end{center}
\caption{Performance quantification (area under the ROC curve) of different methods with respect to their abilities to reduce the 3 types of false connections and their abilities to estimate the true positive connections using the simulated LFP dataset.}
\label{lfpQuant}
\end{figure}

For population recordings, our methods have similar performance to the thalamocortical model example. While the correlation-based methods still suffering from the problem of type 1 false connections (Fig.~\ref{lfpOld}), our differential covariance-based methods can reduce all 3 types of false connections (Fig.~\ref{lfpNew}).  In Fig.~\ref{lfpQuant}, each estimator's performance on LFP data is quantified. In this example, with sufficient data samples, our differential covariance-based methods achieve similar performance to the GLM method. However, for smaller sample sizes, our new methods perform better than the GLM method.

%\FloatBarrier
\subsection{Simulated calcium imaging results}

\begin{figure}[hbtp]
\begin{center}
\includegraphics[width=\textwidth]{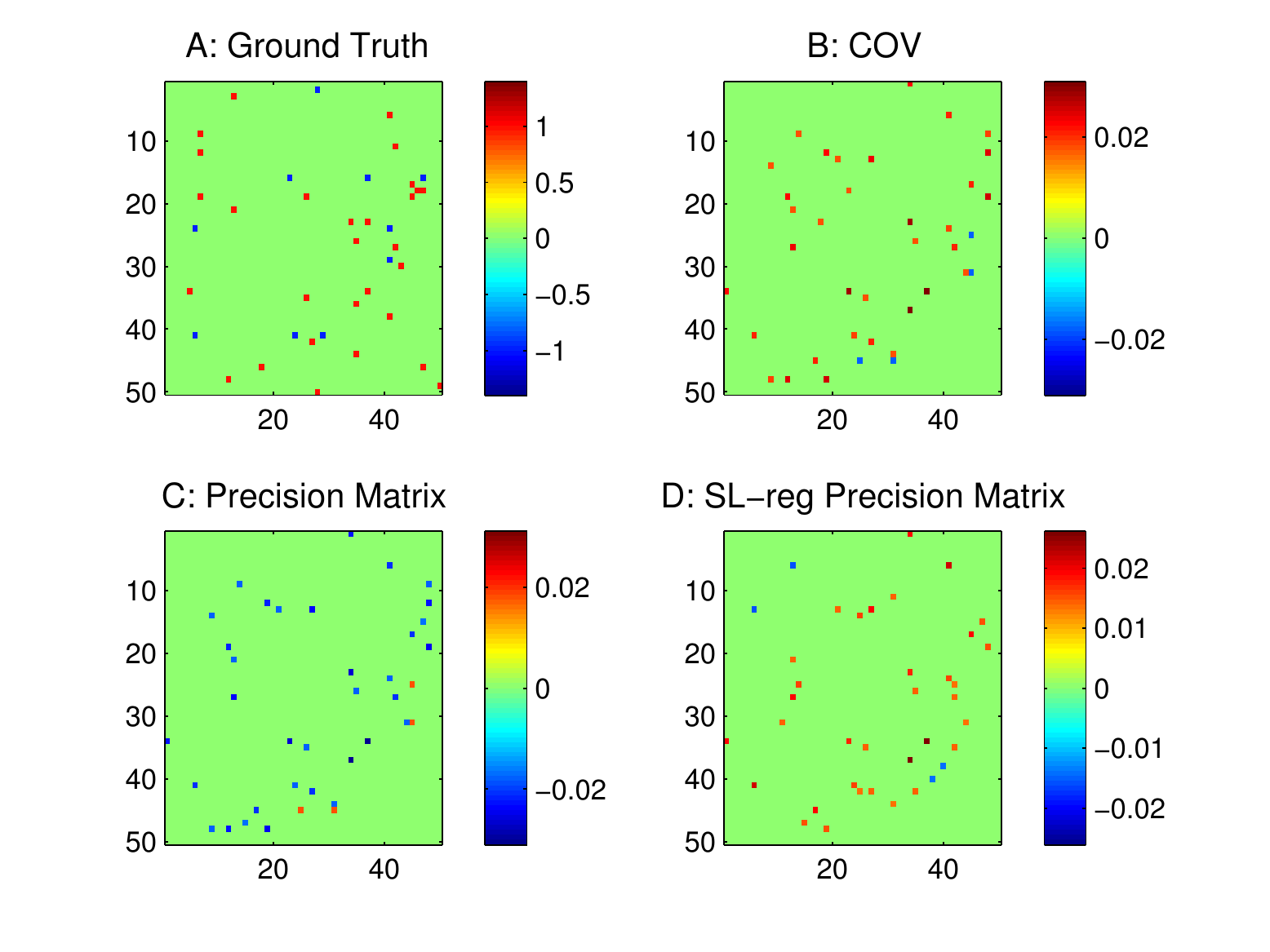}
\end{center}
\caption{Analysis of the simulated calcium imaging dataset with correlation-based methods. A) Ground truth connection matrix B) Estimation from the correlation method. C) Estimation from the precision matrix method. D) Estimation from the sparse+latent regularized precision matrix method. For clarity purpose, panel B,C,D are thresholded to show only the most strong connections, so one can compare it with the ground truth.}
\label{CaOld}
\end{figure}

\begin{figure}[!htb]
\begin{center}
\includegraphics[width=\textwidth]{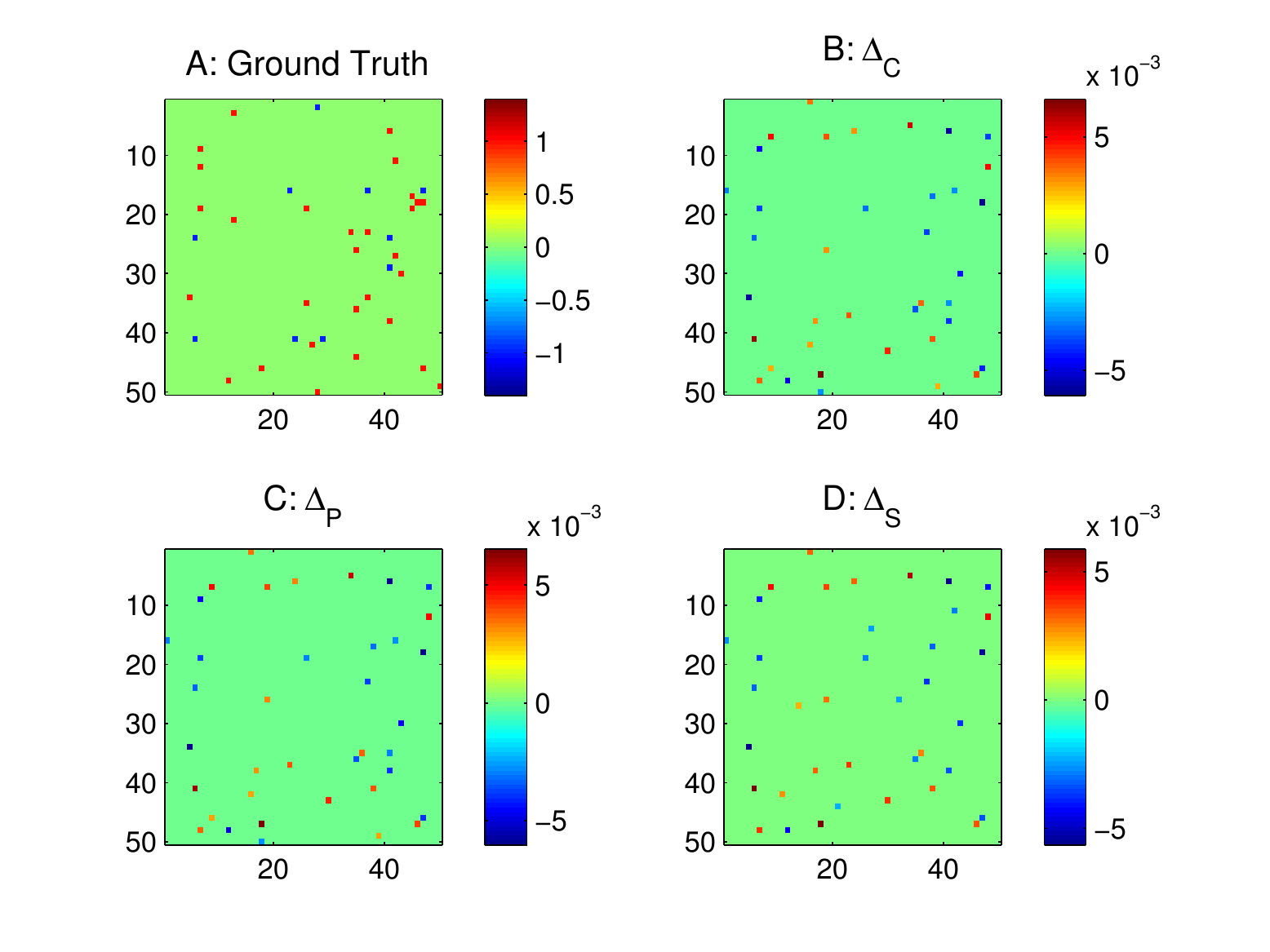}
\end{center}
\caption{Analysis of the simulated calcium imaging dataset with differential covariance-based methods. The color in B,C,D indicates direction of the connections. For element $A_{ij}$, warm color indicates $i$ is the sink, $j$ is the source, i.e. $i \leftarrow j$, and cool color indicates $j$ is the sink, $i$ is the source,  i.e. $i \rightarrow j$. A) Ground truth connection matrix. B) Estimation from the differential covariance method. C) Estimation from the  partial differential covariance method. D) Estimation from the sparse+latent regularized partial differential covariance method. For clarity purpose, panel B,C,D are thresholded to show only the most strong connections, so one can compare it with the ground truth.}
\label{CaNew}
\end{figure}

\begin{figure}[!htb]
\begin{center}
\includegraphics[width=\textwidth]{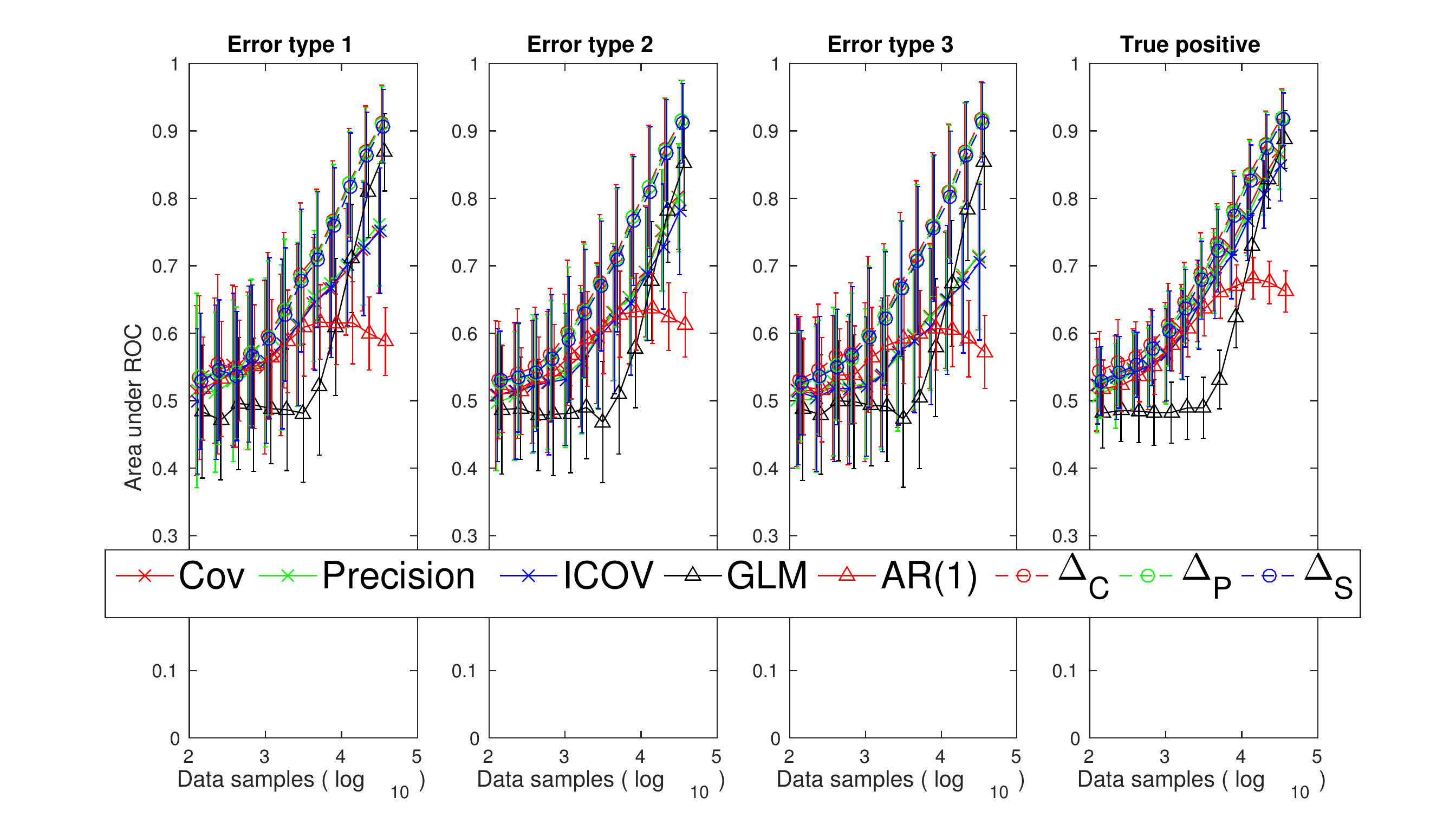}
\end{center}
\caption{Performance quantification (area under the ROC curve) of different methods with respect to their abilities to reduce the 3 types of false connections and their abilities to estimate the true positive connections using the simulated calcium imaging dataset. Error bar is the standard deviation across 100 sets of experiments. Each experiment randomly recorded 50 neurons in a large network. The markers on the plots indicate the average area under the ROC curve values across the 100 sets of experiments. }
\label{CaQuant}
\end{figure}

Lastly, because current techniques only allow recording of a small percentage of neurons in the brain,
we tested our methods on a calcium imaging dataset of 50 neurons recorded from 1000 neurons networks.
In this example, our differential covariance-based methods (Fig.~\ref{CaNew}) match better with the ground truth than the correlation-based methods (Fig.~\ref{CaOld}). 

In Fig.~\ref{CaQuant}, We performed 25 sets of recordings with 50 neurons randomly selected in each of the 4 large networks and quantified the results. 
The markers on the plots are the average area under the ROC curve values across the 100 sets, and
the error bars indicate the standard deviations across these 100 sets of recordings.
Our differential covariance-based methods perform better than the GLM method, and the performance differences seem to be greater in situations with fewer data samples.

%\FloatBarrier
\section{Discussion}
\label{sec_discussion}

\subsection{Generalizability and applicability of the differential covariance-based methods to real experimental data}

Many methods have been proposed to solve the problem of reconstructing the connectivity of a neural network.
\cite{winterhalder2005comparison} reviewed the non-parametric methods and Granger causality based methods. 
Many progress have been made recently using kinetic Ising model and Hopfield network \citep{huang2013sparse,dunn2013learning,battistin2015belief,capone2015inferring,roudi2011mean} with sparsity regularization \citep{pernice2013reconstruction}.
The GLM method \citep{okatan2005analyzing,truccolo2005point, pillow2008spatio} and the maximum entropy method \citep{schneidman2006weak} are two popular classes of methods, which are the main modern approaches for modeling multi-unit recordings \citep{roudi2015multi}.

In the trend of current research, people are recording more and more neurons and looking for new data analysis techniques to handle bigger data with higher dimensionality. 
The field is in favor of algorithms that require fewer samples and scale well with dimensionality, but at the same time not sacrificing the accuracy. 
Also, an algorithm that is model free or make minimum assumptions about the hidden structure of the data has the potential to be applied to multiple types of neural recordings.

The key difference between our methods and other methods is that we use the relationship between a neuron's differential voltage and a neuron's voltage rather than finding the relationship between voltages. 
This provides better performance because the differential voltage is a proxy of a neuron's synaptic current. 
And the relationship between a neuron's synaptic current and its input voltages is more linear, which is suitable for data analysis techniques like the covariance method. 
While this linear relationship hold only for our passive neuron model,
we still see similar or better performance of our methods in our Hodgkin-Huxley model based examples, where we relaxed this assumption and allowed loops in the networks. 
This implies that this class of methods are still applicable even when the ion channels' conductances vary non-linearly with the voltages, which makes the linear relationship only weakly holds.  

%\subsection{Relationship with autoregressive model}
%Detailed discussion of the multivariate autoregressive model and its estimators has been discussed in \cite{harrison2003multivariate}.
%In discrete space, our differential covariance estimator is similar to the MSE estimator of a regularized AR(2) model. 
%%Autoregressive model (AR) has been used in neuroscience data analysis and partially under the name effective connectivity (mcintosh and lima 1994, Friston2011 review, Buchel and Friston). While the general high-order multivariate autoregfressive model has been discussed before (harrison friston 2003), the following regularized AR(2) model is new.
%Following the definition in Eq.~\ref{passiveModel}, we have:
%\begin{equation}
%\begin{array}{c}
%C\frac{dV(t)}{dt} =G \cdot V(t) + \mathcal{N}
%\end{array}
%\end{equation}
%where, $V(t)$ are neurons' membrane voltages. $G$ is the connection matrix that describes the conductance between each pair of neurons. $\mathcal{N}$ is the Gaussian noise.
%
%We note here:
%\begin{equation}
%\begin{array}{c}
%C\frac{dV(t)}{dt} = \frac{V(t+1)-V(t-1)}{2\Delta t}=G \cdot V(t) + \mathcal{N}	\\
%V(t+1)=2\Delta t G \cdot V(t)+V(t-1)+ 2\Delta t \mathcal{N}
%\end{array}
%\end{equation}
%The MSE estimator of $G$ is $\frac{(V(t+1)-V(t-1))V(t)^T}{V(t)V(t)^T}$, where the numerator is our differential covariance estimator. Therefore, the model we proposed is equivalent to an AR(2) model. However, the transition matrix of the $2^{nd}$ order is restricted to be an identity matrix.

\subsection{Caveats and future directions}

One open question for the differential covariance-based methods is how to improve the way they handle neural signals that are non-linearly transformed from the intracellular voltages. 
Currently, to achieve good performance in the calcium imaging example, we need to assume knowing the exact transfer function and reversely reconstruct the action potentials. We find this reverse transform method to be prone to additive Gaussian noise. 
Further study is needed to find better way to preprocess calcium imaging data for the differential covariance-based methods. 

Throughout our simulations, the differential covariance method has better performance and need fewer data samples in some simulations, but not in other simulations. Further investigation is needed to understand where that improvement comes from.

Our Hodgkin-Huxley simulations did not include axonal or synaptic delays, which is a critical feature of a real neural circuit. Unfortunately, it is non-trivial to add this feature to our Hodgkin-Huxley model. Nevertheless, we tested our methods with the passive neuron model using the same connection patterns but with random synaptic delays between neurons. In appendix~\ref{suppFig}, we show that for up to a 10 ms uniformly distributed synaptic delay pattern, our methods still outperform the correlation-based methods.

\section*{Acknowledgement}
We would like to thank Dr. Thomas Liu, and all members of the Computational Neurobiology Lab for providing helpful feedback. This research is supported by ONR MURI (N000141310672), Office of Naval Research, MURI N00014-13-1-0205, Swartz Foundation and Howard Hughes Medical Institute.

\section*{Appendix}
\appendix
\section{Differential covariance derivations}
\label{diffCovMath}

In this section, we first build a simple 3-neuron network to demonstrate that our differential covariance-based methods can reduce the type 1 false connections. Then we develop a generalized theory, which shows that the type 1 false connections' strength is always lower in our differential covariance-based methods than the original correlation-based methods. 

\subsection{A 3-neuron network}
\label{3neurons}

Let us assume a network of 3 neurons, where neuron A projects to neuron B and C:
\begin{equation}
\begin{array}{c}
I_A=dV_A/dt=g_lV_A+\mathcal{N}_A\\\
I_B=dV_B/dt=g_1V_A + g_lV_B+\mathcal{N}_B\\\
I_C=dV_C/dt=g_2V_A + g_lV_C+\mathcal{N}_C
\end{array}
\label{3neurons}
\end{equation}
Here, the cell conductance is $g_l$, neuron A's synaptic connection strength to neuron B is $g_1$, and neuron A's synaptic connection strength to neuron C is $g_2$. $\mathcal{N}_A, \mathcal{N}_B, \mathcal{N}_C$ are independent white Gaussian noises.

From Eq.18 of \cite{fan2011covariances}, we can derive the covariance matrix of this network:
\begin{equation}
\begin{array}{c}
vec(COV) = -(G\otimes I_n+I_{n}\otimes G)^{-1}(D\otimes D)vec(I_m)
\end{array}
\end{equation}
Where,
\begin{equation}
G= \left[ \begin{array}{ccc}
g_l & g_1& g_2\\
0 & g_l& 0\\
0 & 0& g_l
\end{array} \right]^T
\end{equation}
is the transpose of the ground truth connection of the network. And,
\begin{equation}
D= \left[ \begin{array}{ccc}
1 & 0& 0\\
0 & 1& 0\\
0 & 0& 1
\end{array} \right]
\end{equation}
since each neuron receives independent noise.
$I_n$ is an identity matrix of the size of G and $I_m$ is an identity matrix of the size of D. $\otimes$ is the Kronecker product and $vec()$ is the column vectorization function.

Therefore, we have the covariance matrix of the network as:
\begin{equation}
COV= \left[ \begin{array}{ccc}
-1/(2*g_l)&g_1/(4*g_l^2)&g_2/(4*g_l^2)\\
g_1/(4*g_l^2)& -(g_1^2 + 2*g_l^2)/(4*g_l^3)& -(g_1*g_2)/(4*g_l^3)\\
g_2/(4*g_l^2)& -(g_1*g_2)/(4*g_l^3)& -(g_2^2 + 2*g_l^2)/(4*g_l^3)
\end{array} \right]
\label{3neuronsCOV}
\end{equation}

When computing the differential covariance, we plug in Eq.~\ref{3neurons}. For example:
\begin{equation}
\begin{array}{c}
COV(I_C, V_B)=g_2COV(V_A,V_B)+g_lCOV(V_C,V_B)
\end{array}
\end{equation}

Therefore, from Eq.~\ref{3neuronsCOV}, we can compute the differential covariance as:
\begin{equation}
\Delta_P= \left[ \begin{array}{ccc}
-1/2& g_1/(4*g_l)& g_2/(4*g_l)\\
-g_1/(4*g_l)&       -1/2&           0\\
-g_2/(4*g_l)&           0&        -1/2
\end{array} \right]
\label{3neuronsDiffCov}
\end{equation}

Notice that, because the ratio between $COV(V_A,V_B)$ and $COV(V_C,V_B)$ is $-g_l/g_2$, in differential covariance, the type 1 false connection $COV(I_C, V_B)$ has value 0.

\subsection{Type 1 false connection's strength is reduced in differential covariance}
\label{type1App}

In this section, we propose a theory. Given a network, which consists of passive neurons in the following form:
\begin{equation}
I_i(t)=C\frac{dV_i(t)}{dt}=\sum\limits_{k\in\{pre_i\}}g_{k \rightarrow i}V_{k}(t) + g_lV_i(t)+dB_i(t)
\end{equation}
Where $ \{ pre_i \}$ is the set of neurons that project to neuron $i$, $g_{k \rightarrow i}$ is the synaptic conductance for the projection from neuron $k$ to neuron $i$. $B_i(t)$ is a Brownian motion.

And further assume that:
\begin{itemize}
  \item All neurons' leakage conductance $g_l$ and membrane capacitance $C$ are constants and the same.
  \item  There is no loop in the network.
  \item $g_{syn} << g_l$, where $g_{syn}$ is the maximum of $|g_{i \rightarrow j}|$, for $\forall i,j$
\end{itemize}

Then, we prove below that: 
\begin{itemize}
  \item  For two neurons that have physical connection, their covariance is $O(\frac{g_{syn}}{g_l^2})$.
  \item For two neurons that do not have physical connection, their covariance is $O(\frac{g_{syn}^2}{g_l^3})$.
  \item  For two neurons that have physical connection, their differential covariance is $O(\frac{g_{syn}}{g_l})$.
  \item For two neurons that do not have physical connection, their differential covariance is $O(\frac{g_{syn}^3}{g_l^3})$.
	\item The type 1 false connection's strength is reduced in differential covariance.
\end{itemize}

%Therefore, 
%\begin{itemize}
% \item In the correlation method, the strength of a non-physical connection is $\frac{g_{syn}}{g_l}$ times that of a physical connection.
% \item In the differential covariance method, the strength of a non-physical connection is $\frac{g_{syn}^2}{g_l^2}$  times that of a physical connection.
%\end{itemize}
%
%Because $g_{syn} << g_l$, the relative strength of the non-physical connections is reduced in the differential covariance method.

\paragraph{Lemma 1}
The asymptotic auto-covariance of a neuron is 

\begin{equation}
Cov[V_i,V_i]=-(1+2\sum\limits_{k \in \{pre_i\}}g_{k \rightarrow i}Cov[V_k,V_i])/2g_l 
\end{equation}
\paragraph{Proof}
From Eq.9 of \cite{fan2011covariances}, we have,
\begin{equation}
d(V_i-E[V_i])=\sum\limits_{k \in \{pre_i\}}g_{k \rightarrow i}(V_k-E[V_k])dt + g_l(V_i-E[V_i])+dB_i(t)
\label{exp}
\end{equation}
Where $E[]$ is the expectation operation. $(t)$ is dropped from $V_i (t)$ when the meaning is unambiguous.

From Theorem 2 of \cite{fan2011covariances}, integrating by parts using It\^o calculus gives
\begin{equation}
\begin{array}{lc}
d((V_i - E[V_i])(V_i-E[V_i]))=d(V_i-E[V_i])\cdot(V_i-E[V_i])\\
+(V_i-E[V_i])\cdot d(V_i-E[V_i])+d[V_i-E[V_i],V_i-E[V_i]]
\end{array}
\end{equation}
Taking the expectation of both sides with Eq.~\ref{exp} gives
\begin{equation}
dCov[V_i,V_i]=2(\sum\limits_{k \in \{pre_i\}}g_{k \rightarrow i}Cov[V_k,V_i]+g_lCov[V_i,V_i])dt+dE[[B_i(t),B_i(t)]]
\label{dCov}
\end{equation}
when $t\rightarrow +\infty$, Eq.~\ref{dCov} becomes
\begin{equation}
\begin{array}{lc}
0= 2(\sum\limits_{k \in \{pre_i\}}g_{k \rightarrow i}Cov[V_k(+\infty),V_i(+\infty)]+g_lCov[V_i(+\infty),V_i(+\infty)])+1\\
Cov[V_i(+\infty),V_i(+\infty)]=-(1+2\sum\limits_{k \in \{pre_i\}}g_{k \rightarrow i}Cov[V_k(+\infty),V_i(+\infty)])/2g_l
\end{array}
\end{equation}

\paragraph{Lemma 2}
The asymptotic covariance between two neurons is 

\begin{equation}
Cov[V_i,V_j]=-(\sum\limits_{k\in{\{pre_i\}}}g_{k \rightarrow i}Cov[V_k,V_j]+\sum\limits_{k\in{\{pre_j\}}}g_{k \rightarrow j}Cov[V_k,V_i])/2g_l
\end{equation}

\paragraph{Proof}
From Eq.9 of \cite{fan2011covariances}, we have,
\begin{equation}
d(V_i-E[V_i])=\sum\limits_{k \in \{pre_i\}}g_{k \rightarrow i}(V_k-E[V_k])dt + g_l(V_i-E[V_i])+dB_i(t)
\label{exp_ij}
\end{equation}
From Theorem 2 of \cite{fan2011covariances}, integrating by parts using It\^o calculus gives
\begin{equation}
\begin{array}{lc}
d((V_i - E[V_i])(V_j-E[V_j]))=d(V_i-E[V_i])\cdot(V_j-E[V_j])\\
+(V_j-E[V_j])\cdot d(V_i-E[V_i])+d[V_i-E[V_i],V_j-E[V_j]]
\end{array}
\end{equation}
Taking the expectation of both sides with Eq.~\ref{exp_ij} gives
\begin{equation}
\begin{array}{lc}
dCov[V_i,V_i]=(\sum\limits_{k \in \{pre_i\}}g_{k \rightarrow i}Cov[V_k,V_i]
+\sum\limits_{k \in \{pre_j\}}g_{k \rightarrow j}Cov[V_k,V_j]\\
+2g_lCov[V_i,V_j])dt+dE[[B_i(t),B_j(t)]]
\end{array}
\label{dCov_ij}
\end{equation}
when $t\rightarrow +\infty$, Eq.~\ref{dCov_ij} becomes

\begin{align}
&\begin{aligned}
0 &= \sum\limits_{k \in \{pre_i\}}g_{k \rightarrow i}Cov[V_k(+\infty),V_i(+\infty)] \\
&\qquad+\sum\limits_{k \in \{pre_j\}}g_{k \rightarrow j}Cov[V_k(+\infty),V_j(+\infty)]\\
&\qquad+2g_lCov[V_i(+\infty),V_j(+\infty)]\\
Cov[V_i(+\infty),V_j(+\infty)] &=-(\sum\limits_{k \in \{pre_i\}}g_{k \rightarrow i}Cov[V_k(+\infty),V_i(+\infty) \\
&\qquad+\sum\limits_{k \in \{pre_j\}}g_{k \rightarrow j}Cov[V_k(+\infty),V_j(+\infty)])/2g_l
 \end{aligned}
\end{align}

\paragraph{Theorem 1}
The auto-covariance of a neuron is $O(\frac{1}{g_l})$.
The covariance of two different neurons with or without physical connection is $O(\frac{g_{syn}}{g_l^2})$.

\paragraph{Proof}
We prove this by induction.
\paragraph{The basis:}
The base case contains two neurons:
\begin{equation}
\begin{array}{lc}
C\frac{dV_1}{dt}= g_lV_1+\mathcal{N}_1\\
C\frac{dV_2}{dt}=g_{1 \rightarrow 2}V_1 + g_lV_2+\mathcal{N}_2
\end{array}
\end{equation}
From Lemma 1, we have:
\begin{equation}
Cov[V_1,V_1]=-1/2g_l
\end{equation}
Then, from Lemma 2, we have:
\begin{equation}
\begin{array}{lc}
Cov[V_1,V_2]=-g_{1 \rightarrow 2}Cov[V_1,V_1]/2g_l\\
Cov[V_1,V_2]=g_{1 \rightarrow 2}/4g_l^2
\end{array}
\end{equation}

And, from Lemma 1, we have:
\begin{equation}
\begin{array}{lc}
Cov[V_2,V_2]=-(1+2g_{1 \rightarrow 2}Cov[V_1,V_2])/2g_l\\
Cov[V_1,V_2]=-1/2g_l-g_{1 \rightarrow 2}^2/4g_l^3
\end{array}
\end{equation}

So the statement holds.
\paragraph{The inductive step:}
If the statement holds for a network of $n-1$ neurons, we add one more neuron to it.

\paragraph{Part 1:}
First, let's prove the covariance of any neuron with n is also $O(\frac{g_{syn}}{g_l^2})$.

From Lemma 2, we have:
\begin{equation}
Cov[V_i,V_n]=-(\sum\limits_{k \in \{pre_i\}}g_{k \rightarrow i}Cov[V_k,V_n]+\sum\limits_{k \in \{pre_n\}}g_{k \rightarrow n}Cov[V_{k},V_i])/2g_l\\
\label{cov_in}
\end{equation}
where, $\{pre_i\}$ are the neurons projecting to neuron $i$, $\{pre_n\}$ are the neurons projecting to neuron $n$.

Note that, because $\{pre_n\}$ are neurons from the old network, $Cov[V_{k},V_i],k \in \{pre_n\}$ is at most $O(\frac{1}{g_l})$, and it is $O(\frac{1}{g_l})$ only when $k=i$.

Now, we need to prove that $Cov[V_{k},V_n],k \in \{pre_i\}$ is also $O(\frac{1}{g_l})$. We prove this by contradiction. Let's suppose that $Cov[V_{k},V_n],k \in \{pre_i\}$ is larger than $O(\frac{1}{g_l})$. Then similar to Eq.~\ref{cov_in}, we have:

For $p \in \{pre_i\}$
\begin{equation}
Cov[V_p,V_n]=-(\sum\limits_{k \in \{pre_p\}}g_{k \rightarrow p}Cov[V_k,V_n]+\sum\limits_{k \in \{pre_n\}}g_{k \rightarrow n}Cov[V_k,V_p])/2g_l\\
\label{cov_pn}
\end{equation}

Here we separate the problem into two situations,

\subparagraph{Case 1: neuron i projects to neuron n}

Since there is no loop in the network, $n \notin \{pre_i\}$. Therefore, $Cov[V_k,V_p], k\in\{pre_n\}, p\in\{pre_i\}$ is the covariance of two neurons from the old network and is $O(\frac{1}{g_l})$. $Cov[V_{k},V_n], k \in \{pre_p\}$ must be larger than $O(\frac{1}{g_l})$, such that 
$Cov[V_p,V_n], p \in \{pre_i\}$ is larger than $O(\frac{1}{g_l})$.

Therefore, if a neuron's covariance with neuron n is larger than $O(\frac{1}{g_l})$, one of its antecedents' covariance with neuron n is also larger than $O(\frac{1}{g_l})$. Since we assume there is no loop in this network, there must be at least one antecedent (say, neuron m) whose covariance with neuron n is larger than $O(\frac{1}{g_l})$ and it has no antecedent.

However, from Lemma 2:
\begin{equation}
Cov[V_m,V_n]=-(\sum\limits_{k \in \{pre_n\}}g_{k \rightarrow n}Cov[V_{k},V_m])/2g_l
\label{cov_mn}
\end{equation}
Since $Cov[V_m,V_k]), k \in \{pre_n\}$ is $O(\frac{1}{g_l})$, $Cov[V_m,V_n]$ is $O(\frac{g_{syn}}{g_l^2})$, which is smaller than $O(\frac{1}{g_l})$. This is a contradiction. So $Cov[V_{k},V_n],k \in \{pre_i\}$ is no larger than $O(\frac{1}{g_l})$. Therefore, $Cov[V_i,V_n]$ is $O(\frac{g_{syn}}{g_l^2})$.

\subparagraph{Case 2: neuron i does not project to neuron n}
Now, in Eq.~\ref{cov_pn}, it is possible that $n \in \{pre_i\}$. However, in case 1, we just proved that the covariance of any neuron that projects to neuron n is $O(\frac{g_{syn}}{g_l^2})$. Therefore, $Cov[V_k,V_p], k \in \{pre_n\}, p \in \{pre_i\}$ is $O(\frac{g_{syn}}{g_l^2})$ regardless of whether $p=n$.

Then, similar to case 1, there must be an antecedent of neuron i (say neuron m), whose covariance with neuron n is larger than $O(\frac{1}{g_l})$ and it has no antecedent. Then from Eq.~\ref{cov_mn} we know this is a contradiction. So, $Cov[V_i,V_n]$ is $O(\frac{g_{syn}}{g_l^2})$.

\paragraph{Part 2:}
Then, for the auto-covariance of neuron n:
\begin{equation}
Cov[V_n,V_n]=-(1+2\sum\limits_{k \in \{pre_n\}}g_{k \rightarrow n}Cov[V_k,V_n])/2g_l 
\label{autocov_n}
\end{equation}

As we already proved that any neuron's covariance with neuron n is $O(\frac{g_{syn}}{g_l^2})$, the dominant term in Eq.~\ref{autocov_n} is $-1/2g_l$. Therefore, the auto-covariance of neuron n is also $O(\frac{1}{g_l})$.

End of proof.

\paragraph{Theorem 2}
The covariance of two neurons that are physically connected is $O(\frac{g_{syn}}{g_l^2})$.
The covariance of two neurons that are not physically connected is $O(\frac{g_{syn}^2}{g_l^3})$.

\paragraph{Proof}
From Lemma 2, we have:
\begin{equation}
Cov[V_i,V_j]=-(\sum\limits_{k \in \{pre_i\}}g_{k \rightarrow i}Cov[V_k,V_j]+\sum\limits_{k \in \{pre_j\}}g_{k \rightarrow j}Cov[V_k,V_i])/2g_l\\
\end{equation}

If neuron $i$ and neuron $j$ are physically connected, let's say $i \rightarrow j$, then $i \in \{pre_j\}$. Thus one of the $V_k$ for $k \in \{pre_j\}$ is $V_i$. Therefore, $Cov[V_k,V_i]$ is $O(\frac{1}{g_l})$. Since there is no loop in the network, $j \notin \{pre_i\}$, so $Cov[V_k,V_j]$ is $O(\frac{g_{syn}}{g_l^2})$. Therefore, $Cov[V_i,V_j]$ is $O(\frac{g_{syn}}{g_l^2})$.

If neuron $i$ and neuron $j$ are not physically connected, we have $i \notin \{pre_j\}$, so $Cov[V_k,V_i])$ is $O(\frac{g_{syn}}{g_l^2})$. And $j \notin \{pre_i\}$, so $Cov[V_k,V_j])$ is $O(\frac{g_{syn}}{g_l^2})$. Therefore, $Cov[V_i,V_j]$ is $O(\frac{g_{syn}^2}{g_l^3})$.

End of proof.

\paragraph{Lemma 3}
The differential covariance of two neurons,
\begin{equation}
Cov[\frac{dV_i}{dt},V_j]+Cov[V_i,\frac{dV_j}{dt}]=0
\end{equation}

\paragraph{Proof}
From Lemma 2, we have,
\begin{equation}
\begin{array}{lc}
2g_lCov[V_i,V_j]+\sum\limits_{k\in{\{pre_i\}}}g_{k \rightarrow i}Cov[V_k,V_j]+\sum\limits_{k\in{\{pre_j\}}}g_{k \rightarrow j}Cov[V_k,V_i]=0 \\
Cov[\sum\limits_{k\in{\{pre_i\}}}g_{k \rightarrow i}V_k+g_lV_i,V_j]+Cov[\sum\limits_{k\in{\{pre_j\}}}g_{k \rightarrow j}V_k+g_lV_j,V_i]=0\\
Cov[\frac{dV_i}{dt},V_j]+Cov[V_i,\frac{dV_j}{dt}]=0
\end{array}
\end{equation}

End of proof.

\paragraph{Theorem 3}
The differential covariance of two neurons that are physically connected is $O(\frac{g_{syn}}{g_l})$.

\paragraph{Proof}
Assume two neurons have physical connection as $i \rightarrow j$.
The differential covariance of them is:
\begin{equation}
\begin{array}{lc}
Cov[\frac{dV_i}{dt},V_j]=Cov[\sum\limits_{k \in \{pre_i\}}g_{k \rightarrow i}V_k+g_lV_i,V_j]\\
=\sum\limits_{k \in \{pre_i\}}g_{k \rightarrow i}Cov[V_k,V_j]+g_lCov[V_i,V_j]
\end{array}
\end{equation}

From Theorem 2, we know $Cov[V_i,V_j]$ is $O(\frac{g_{syn}}{g_l^2})$, and
\begin{itemize}
\item If $V_k$ is projecting to $V_j$, $Cov[V_k,V_j]$ is $O(\frac{g_{syn}}{g_l^2})$.
\item If $V_k$ is not projecting to $V_j$, $Cov[V_k,V_j]$ is $O(\frac{g_{syn}^2}{g_l^3})$.
\end{itemize}
Therefore, the dominant term is $Cov[V_i,V_j]$. And $Cov[\frac{dV_i}{dt},V_j]$ is $O(\frac{g_{syn}}{g_l})$.

From Lemma 3,
\begin{equation}
Cov[V_i,\frac{dV_j}{dt}]=-Cov[\frac{dV_i}{dt},V_j]
\end{equation}
Therefore, $Cov[V_i,\frac{dV_j}{dt}]$ is $O(\frac{g_{syn}}{g_l})$.

%The other differential covariance of them is:
%\begin{equation}
%\begin{array}{lc}
%Cov[V_i, \frac{dV_j}{dt}]=Cov[V_i, \sum\limits_{k \in pre_j}g_{syn_k}^{pre_j}V_k+g_lV_j]\\
%=\sum\limits_{k \in pre_j}g_{syn_k}^{pre_j}Cov[V_i,V_k]+g_lCov[V_i,V_j]
%\end{array}
%\end{equation}
%
%From Theorem 2, we know $Cov[V_i,V_j]$ is $O(\frac{g_{syn}}{g_l^2})$. For $Cov[V_i,V_k]$, there are 3 possibilities.
%\begin{itemize}
%\item if $V_k$ is $V_i$, $Cov[V_i,V_k]$ is $O(\frac{1}{g_l})$.
%\item if $V_k$ is projecting to $V_i$, $Cov[V_i,V_k]$ is $O(\frac{g_{syn}}{g_l^2})$.
%\item if $V_k$ is not projecting to $V_i$, $Cov[V_i,V_k]$ is $O(\frac{g_{syn}^2}{g_l^3})$.
%\end{itemize}
%
%So, $Cov[V_i,V_k]$ is at most $O(\frac{1}{g_l})$. Therefore, $Cov[V_i, \frac{dV_j}{dt}]$ is $O(\frac{G_{syn}}{g_l})$.

End of proof.

\begin{figure}[hbtp]
\begin{center}
\includegraphics[width=\textwidth]{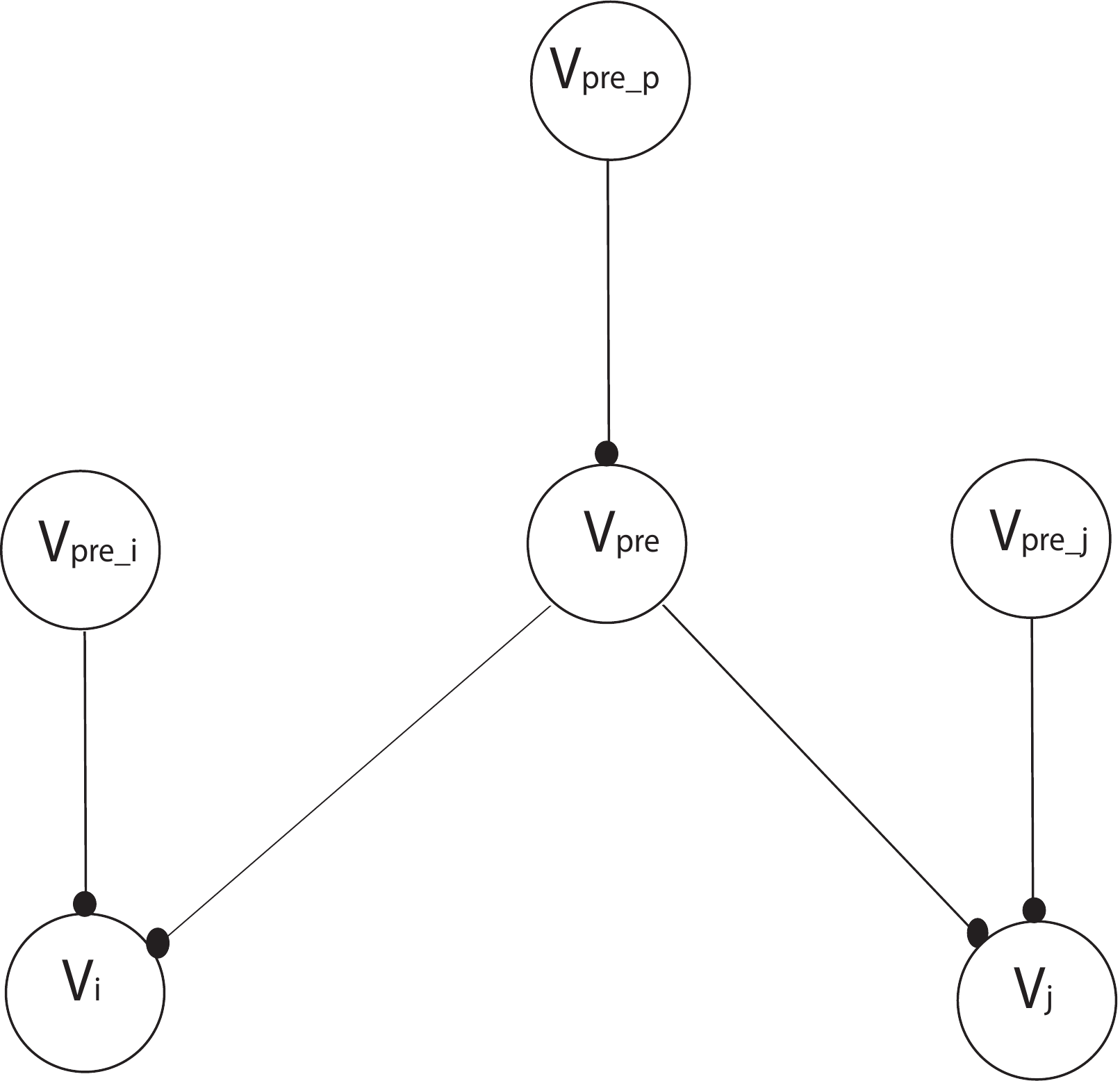}
\end{center}
\caption{The network used in Theorem 4's proof.}
\label{ij_antecedents}
\end{figure}

\FloatBarrier
\paragraph{Theorem 4}
The differential covariance of two neurons that are not physically connected is $O(\frac{g_{syn}^3}{g_l^3})$.

\paragraph{Proof}
First, let's define the antecedents of neuron i and neuron j.
Shown in Fig.~\ref{ij_antecedents},
$\{pre\}$ is the set of common antecedents of neuron $i$ and neuron $j$.
$\{pre_p\}$ is the set of antecedents of $p \in \{pre\}$.
$\{pre_i\}$ is the set of exclusive antecedents of neuron $i$.
$\{pre_j\}$ is the set of exclusive antecedents of neuron $j$.

From Lemma 2, we have, for any neuron $p \in {\{pre\}}$
\begin{equation}
\begin{split}
Cov[V_i,V_{p}]=-(\sum\limits_{k\in{\{pre_i\}}}g_{k \rightarrow i}Cov[V_{k},V_{p}]
 +\sum\limits_{k\in{\{pre\}}}g_{k \rightarrow i}Cov[V_k,V_{p}] \\
 +\sum\limits_{k\in{\{pre_p\}}}g_{k \rightarrow p}Cov[V_k,V_i]
)/2g_l
\end{split}
\label{cov_ip}
\end{equation}

\begin{equation}
\begin{split}
Cov[V_j,V_{p}]=-(\sum\limits_{k\in{\{pre_j\}}}g_{k \rightarrow j}Cov[V_{k},V_{p}]
 +\sum\limits_{k\in{\{pre\}}}g_{k \rightarrow j}Cov[V_k,V_{p}] \\
 +\sum\limits_{k\in{\{pre_p\}}}g_{k \rightarrow p}Cov[V_k,V_j]
)/2g_l
\end{split}
\label{cov_jp}
\end{equation}

For simplicity, we define
\begin{equation}
\begin{array}{lc}
\sum\limits_{k\in{\{pre_j\}}}g_{k \rightarrow j}Cov[V_{k},V_{p}] = C_p\\
\sum\limits_{k\in{\{pre_i\}}}g_{k \rightarrow i}Cov[V_{k},V_{p}] = D_p \\
\sum\limits_{k\in{\{pre_p\}}}g_{k \rightarrow p}Cov[V_k,V_j] = E_p \\
\sum\limits_{k\in{\{pre_p\}}}g_{k \rightarrow p}Cov[V_k,V_i] = F_p 
\end{array}
\end{equation}

From Lemma 2, we also have,
\begin{equation}
\begin{split}
Cov[V_i,V_j]=-(\sum\limits_{k\in{\{pre_i\}}}g_{k \rightarrow i}Cov[V_{k},V_j]
 +\sum\limits_{k\in{\{pre_j\}}}g_{k \rightarrow j}Cov[V_{k},V_i] \\
 +\sum\limits_{k\in{\{pre\}}}g_{k \rightarrow i}Cov[V_k,V_j]
+\sum\limits_{k\in{\{pre\}}}g_{k \rightarrow j}Cov[V_k,V_i]
)/2g_l
\end{split}
\label{cov_ij}
\end{equation}

For simplicity, we define
\begin{equation}
\begin{array}{lc}
\sum\limits_{k\in{\{pre_i\}}}g_{k \rightarrow i}Cov[V_{k},V_j] = A\\
\sum\limits_{k\in{\{pre_j\}}}g_{k \rightarrow j}Cov[V_{k},V_i] = B
\end{array}
\end{equation}

Plug in Eq.~\ref{cov_ip}, Eq.~\ref{cov_jp} to Eq.~\ref{cov_ij}, we have,
\begin{equation}
\begin{split}
Cov[V_i,V_j]=\sum\limits_{p\in{\{pre\}}}\frac{g_{p \rightarrow i}}{4g_l^2}(\sum\limits_{k\in{\{pre\}}}g_{k \rightarrow j}Cov[V_k,V_{p}]+C_p+E_p)\\
+\sum\limits_{p\in{\{pre\}}}\frac{g_{p \rightarrow j}}{4g_l^2}(\sum\limits_{k\in{\{pre\}}}g_{k \rightarrow i}Cov[V_k,V_{p}] + D_p+F_p)\\
-\frac{A}{2g_l}-\frac{B}{2g_l}
\end{split}
\label{cov_ijv2}
\end{equation}
Now, we look at the differential covariance between neuron i and neuron j,
\begin{equation}
\begin{array}{lc}
Cov[\frac{dV_i}{dt},V_j]=\sum\limits_{k\in{\{pre_i\}}}g_{k \rightarrow i}Cov[V_{k},V_j]
 +\sum\limits_{k\in{\{pre\}}}g_{k \rightarrow i}Cov[V_k,V_j] + g_lCov[V_i,V_j]
\end{array}
\end{equation}

Plug in Eq.~\ref{cov_jp}, Eq.~\ref{cov_ijv2},  we have,

\begin{align}
&\begin{aligned}
Cov[\frac{dV_i}{dt},V_j]&=A +\sum\limits_{p\in{\{pre\}}}\frac{-g_{p \rightarrow i}}{2g_l}(\sum\limits_{k\in{\{pre\}}}g_{k \rightarrow j}Cov[V_k,V_{p}]+C_p+E_p) \\
&\qquad+ g_l[\sum\limits_{p\in{\{pre\}}}\frac{g_{p \rightarrow i}}{4g_l^2}
(\sum\limits_{k\in{\{pre\}}}g_{k \rightarrow j}Cov[V_k,V_{p}]+C_p+E_p)\\
&\qquad+\sum\limits_{p\in{\{pre\}}}\frac{g_{p \rightarrow j}}{4g_l^2}(\sum\limits_{k\in{\{pre\}}}g_{k \rightarrow i}Cov[V_k,V_{p}] + D_p+F_p)\\
&\qquad-\frac{A}{2g_l}-\frac{B}{2g_l}]\\
&=\frac{A}{2}-\frac{B}{2}+\sum\limits_{p\in{\{pre\}}}\frac{g_{p \rightarrow j}}{4g_l}(D_p+F_p)
-\sum\limits_{p\in{\{pre\}}}\frac{g_{p \rightarrow i}}{4g_l}(C_p+E_p)
 \end{aligned}
\end{align}

Note,
\begin{itemize}
\item There is no physical connection between a neuron in $\{pre_i\}$ and neuron j, otherwise, this neuron belongs to $\{pre\}$. Therefore, from Theorem 2, $A$ is $O(\frac{g_{syn}^2}{g_l^3})*g_{k \rightarrow i}=O(\frac{g_{syn}^3}{g_l^3})$.
\item There is no physical connection between a neuron in $\{pre_j\}$ and neuron i, otherwise, this neuron belongs to $\{pre\}$. Therefore, from Theorem 2, $B$ is $O(\frac{g_{syn}^2}{g_l^3})*g_{k \rightarrow j}=O(\frac{g_{syn}^3}{g_l^3})$.
\item There could be physical connections between neurons in $\{pre_p\}$ and $\{pre_j\}$, so $C_p$ is $O(\frac{g_{syn}}{g_l^2})*g_{k \rightarrow j}=O(\frac{g_{syn}^2}{g_l^2})$.
\item There could be physical connections between neurons in $\{pre_p\}$ and $\{pre_i\}$, so $D_p$ is $O(\frac{g_{syn}}{g_l^2})*g_{k \rightarrow i}=O(\frac{g_{syn}^2}{g_l^2})$.
\item There could be physical connections between neurons in $\{pre_p\}$ and neuron $j$, so $E_p$ is $O(\frac{g_{syn}}{g_l^2})*g_{k \rightarrow p}=O(\frac{g_{syn}^2}{g_l^2})$.
\item There could be physical connections between neurons in $\{pre_p\}$ and neuron $i$, so $F_p$ is $O(\frac{g_{syn}}{g_l^2})*g_{k \rightarrow p}=O(\frac{g_{syn}^2}{g_l^2})$.
\end{itemize}

Therefore, 

\begin{align}
&\begin{aligned}
Cov[\frac{dV_i}{dt},V_j]&= O(\frac{g_{syn}^3}{g_l^3})-O(\frac{g_{syn}^3}{g_l^3})+\sum\limits_{p\in{\{pre\}}}\sum\limits_{k\in{\{pre\}}}\frac{g_{p \rightarrow j}}{4g_l}(O(\frac{g_{syn}^2}{g_l^2})+O(\frac{g_{syn}^2}{g_l^2})) \\
&\qquad-\sum\limits_{p\in{\{pre\}}}\sum\limits_{k\in{\{pre\}}}\frac{g_{p \rightarrow i}}{4g_l}(O(\frac{g_{syn}^2}{g_l^2})+O(\frac{g_{syn}^2}{g_l^2})) \\
&=O(\frac{g_{syn}^3}{g_l^3})
 \end{aligned}
\end{align}

From Lemma 3, we know, $Cov[V_i, \frac{dV_j}{dt}]$ is also $O(\frac{g_{syn}^3}{g_l^3})$.

End of proof.

\paragraph{Theorem 5}
The type 1 false connection's strength is reduced in differential covariance

\paragraph{Proof}
From theorem 1 and theorem 2, we know that,
in the correlation method, the strength of a non-physical connection ($O(\frac{g_{syn}}{g_l^2})$) is $\frac{g_{syn}}{g_l}$ times that of a physical connection ($O(\frac{g_{syn}^2}{g_l^3})$).

From theorem 3 and theorem 4, we know that,
in the differential covariance method, the strength of a non-physical connection ($O(\frac{g_{syn}}{g_l})$) is $\frac{g_{syn}^2}{g_l^2}$  times that of a physical connection ($O(\frac{g_{syn}^3}{g_l^3})$).

Because $g_{syn} << g_l$, the relative strength of the non-physical connections is reduced in the differential covariance method.

End of proof.

\subsection{Directionality information in differential covariance}
\label{directionality}

\paragraph{Theorem 6}
If neuron $i$ projects to neuron $j$ with an excitatory connection, $Cov[\frac{dV_i}{dt},V_j]>0$ and $Cov[V_i, \frac{dV_j}{dt}]<0$

\paragraph{Proof}
Given the model above, similar to Theorem 4, from Lemma 2, we have, for any neuron $p \in {\{pre\}}$
\begin{equation}
\begin{split}
Cov[V_i,V_{p}]=-(\sum\limits_{k\in{\{pre_i\}}}g_{k \rightarrow i}Cov[V_{k},V_{p}]
 +\sum\limits_{k\in{\{pre\}}}g_{k \rightarrow i}Cov[V_k,V_{p}] \\
 +\sum\limits_{k\in{\{pre_p\}}}g_{k \rightarrow p}Cov[V_k,V_i]
)/2g_l
\end{split}
\end{equation}

\begin{equation}
\begin{split}
Cov[V_j,V_{p}]=-(\sum\limits_{k\in{\{pre_j\}}}g_{k \rightarrow j}Cov[V_{k},V_{p}]
 +\sum\limits_{k\in{\{pre\}}}g_{k \rightarrow j}Cov[V_k,V_{p}] \\
 +\sum\limits_{k\in{\{pre_p\}}}g_{k \rightarrow p}Cov[V_k,V_j]+g_{i \rightarrow j}Cov[V_i,V_p]
)/2g_l
\end{split}
\end{equation}

For simplicity, we define
\begin{equation}
\begin{array}{lc}
\sum\limits_{k\in{\{pre_j\}}}g_{k \rightarrow j}Cov[V_{k},V_{p}] = C_p\\
\sum\limits_{k\in{\{pre_i\}}}g_{k \rightarrow i}Cov[V_{k},V_{p}] = D_p \\
\sum\limits_{k\in{\{pre_p\}}}g_{k \rightarrow p}Cov[V_k,V_j] = E_p \\
\sum\limits_{k\in{\{pre_p\}}}g_{k \rightarrow p}Cov[V_k,V_i] = F_p 
\end{array}
\end{equation}

From Lemma 2, we also have,
\begin{equation}
\begin{split}
Cov[V_i,V_j]=-(\sum\limits_{k\in{\{pre_i\}}}g_{k \rightarrow i}Cov[V_{k},V_j]
 +g_{i \rightarrow j}Cov[V_i,V_i]
 +\sum\limits_{k\in{\{pre_j\}}}g_{k \rightarrow j}Cov[V_{k},V_i] \\
 +\sum\limits_{k\in{\{pre\}}}g_{k \rightarrow i}Cov[V_k,V_j]
+\sum\limits_{k\in{\{pre\}}}g_{k \rightarrow j}Cov[V_k,V_i]
)/2g_l
\end{split}
\end{equation}

For simplicity, we define
\begin{equation}
\begin{array}{lc}
\sum\limits_{k\in{\{pre_i\}}}g_{k \rightarrow i}Cov[V_{k},V_j] = A\\
\sum\limits_{k\in{\{pre_j\}}}g_{k \rightarrow j}Cov[V_{k},V_i] = B
\end{array}
\end{equation}

Plug in, we have,
\begin{equation}
\begin{split}
Cov[V_i,V_j]=\sum\limits_{p\in{\{pre\}}}\frac{g_{p \rightarrow i}}{4g_l^2}(\sum\limits_{k\in{\{pre\}}}g_{k \rightarrow j}Cov[V_k,V_{p}]+C_p+E_p+g_{i \rightarrow j}Cov[V_i,V_p])\\
+\sum\limits_{p\in{\{pre\}}}\frac{g_{p \rightarrow j}}{4g_l^2}(\sum\limits_{k\in{\{pre\}}}g_{k \rightarrow i}Cov[V_k,V_p] + D_p+F_p)\\
-\frac{A}{2g_l}-\frac{B}{2g_l}-\frac{g_{i \rightarrow j}Cov[V_i,V_i]}{2g_l}
\end{split}
\end{equation}
Now, we look at the differential covariance between neuron i and neuron j,

\begin{align}
&\begin{aligned}
Cov[\frac{dV_i}{dt},V_j]
&=\frac{A}{2}-\frac{B}{2}+\sum\limits_{p\in{\{pre\}}}\frac{g_{p \rightarrow j}}{4g_l}(D_p+F_p)\\
&-\sum\limits_{p\in{\{pre\}}}\frac{g_{p \rightarrow i}}{4g_l}(C_p+E_p++g_{i \rightarrow j}Cov[V_i,V_p])-\frac{g_{i \rightarrow j}Cov[V_i,V_i]}{2}
\end{aligned}
\end{align}

Note, in Theorem 4, we already proved the scale of $A, B, C_p, D_p, E_p, F_p$. We also have,
\begin{itemize}
\item There are physical connections between neurons in $\{pre\}$ and neuron $i$, so $g_{i \rightarrow j}Cov[V_i,V_p]$ is $O(\frac{g_{syn}}{g_l^2})*g_{k \rightarrow p}=O(\frac{g_{syn}^2}{g_l^2})$.
\item From Lemma 1, the auto-covariance of neuron $i$ is $O(\frac{1}{g_l})$. So $g_{i \rightarrow j}Cov[V_i,V_i]$ is $O(\frac{1}{g_l})*g_{i \rightarrow j}=O(\frac{g_{syn}}{g_l})$.
\end{itemize}

Therefore, $g_{i \rightarrow j}Cov[V_i,V_i]$ is the dominant term in $Cov[\frac{dV_i}{dt},V_j]$. Since $Cov[V_i,V_i]>0$, for excitatory connection $g_{i \rightarrow j}>0$, $Cov[\frac{dV_i}{dt},V_j]<0$.

From Lemma 3,
\begin{equation}
Cov[V_i,\frac{dV_j}{dt}]=-Cov[\frac{dV_i}{dt},V_j]
\end{equation}
Therefore, $Cov[V_i,\frac{dV_j}{dt}]>0$.

End of proof.

\paragraph{Corollary 1}
If neuron $i$ projects to neuron $j$ with an inhibitory connection, $Cov[\frac{dV_i}{dt},V_j]>0$ and $Cov[V_i, \frac{dV_j}{dt}]<0$

\paragraph{Proof}

The proof is similar to Theorem 6. Again, we know $g_{i \rightarrow j}Cov[V_i,V_i]$ is the dominant term in $Cov[\frac{dV_i}{dt},V_j]$. Since $Cov[V_i,V_i]>0$, now for an inhibitory  connection $g_{i \rightarrow j}<0$, $Cov[\frac{dV_i}{dt},V_j]>0$.

From Lemma 3,
\begin{equation}
Cov[V_i,\frac{dV_j}{dt}]=-Cov[\frac{dV_i}{dt},V_j]
\end{equation}
Therefore, $Cov[V_i,\frac{dV_j}{dt}]<0$.

End of proof.

\section{Benchmarked methods}
\label{corrMethod}
We compared our methods to a few popular methods.

\subsection{Covariance method}
The covariance matrix is defined as:
\begin{equation}
	COV_{x,y}=\frac{1}{N}\sum^{N}_{i=1}(x_i-\mu_{x})(y_i-\mu_{y})
\end{equation}
Where $x$ and $y$ are two variables. $\mu_{x}$ and $\mu_{y}$ are their population mean.

\subsection{Precision matrix}
\label{sec_PCOV}
The precision matrix is the inverse of the covariance matrix:
\begin{equation}
P=COV^{-1},
\end{equation}
It can be considered as one kind of partial correlation. Here we briefly review this derivation, because we use it to develop our new method. The derivation here is based on and adapted from \cite{cox1996multivariate}.

We begin by considering a pair of variables $(x, y)$, and remove the correlation in them introduced from a control variable $z$. 

First, we define the covariance matrix as:
\begin{equation}
COV_{xyz}= \left[ \begin{array}{ccc}
\sigma_{xx} & \sigma_{xy}& \sigma_{xz}\\
\sigma_{yx} & \sigma_{yy}& \sigma_{yz}\\
\sigma_{zx} & \sigma_{zy}& \sigma_{zz}
\end{array} \right]
\end{equation}
By solving the linear regression problem:
\begin{equation}
\begin{array}{rcl}
w_x=\argmin\limits_{w}E(x-w*z)^2\\
w_y=\argmin\limits_{w}E(y-w*z)^2
\end{array}
\end{equation}
we have:
\begin{equation}
\begin{array}{rcl}
w_x=\sigma_{xz}\sigma_{zz}^{-1}\\
w_y=\sigma_{yz}\sigma_{zz}^{-1}
\end{array}
\end{equation}

then, we define the residual of $x,y$ as,
\begin{equation}
\begin{array}{c}
r_x = x- w_x*z\\
r_y = y - w_y*z
\end{array}
\end{equation}
Therefore, the covariance of $r_x , r_y$ is:
\begin{equation}
COV_{r_x, r_y}= \sigma_{xy} - \sigma_{xz}*\sigma_{zz}^{-1}*\sigma_{yz}
\end{equation}

On the other hand, if we define the precision matrix as:
\begin{equation}
P_{xyz}= \left[ \begin{array}{ccc}
p_{xx} & p_{xy}& p_{xz}\\
p_{yx} & p_{yy}& p_{yz}\\
p_{zx} & p_{zy}& p_{zz}
\end{array} \right]
\end{equation}

Using Cramer's rule, we have:
 \begin{equation}
p_{xy}= \frac{-\left| \begin{array}{cc}
\sigma_{xy} & \sigma_{xz}\\
\sigma_{zy} & \sigma_{zz}
\end{array} \right|}{|COV_{xyz}|}
\end{equation}
Therefore,
\begin{equation}
p_{xy}= \frac{-\sigma_{zz}}{{|COV_{xyz}|}}(\sigma_{xy} - \sigma_{xz}*\sigma_{zz}^{-1}*\sigma_{yz})
\end{equation}

So $p_{xy}$ and $COV_{r_x,r_y}$ are differed by a ratio of $ \frac{-\sigma_{zz}}{{|COV_{xyz}|}}$.

\subsection{Sparse latent regularization}
\label{SLreg}

Prior studies\citep{banerjee2006convex, friedman2008sparse} have shown that regularizations can provide better estimation if the ground truth connection matrix has a known structure (e.g. sparse). For all data tested in this paper, the sparse latent regularization\citep{yatsenko2015improved} worked best. For a fair comparison, we applied the sparse latent regularization to both the precision matrix method and our differential covariance method.

In the original sparse latent regularization method, people made the assumption that a larger precision matrix $S$ is the joint distribution of the $p$ observed neurons and $d$ latent neurons \citep{yatsenko2015improved}. i.e.

\[S= \left( \begin{array}{cc}
S_{11} & S_{12}\\
S_{21} & S_{22}\end{array} \right)\] 

Where $S_{11}$ corresponds to the observable neurons.
If we can only measure the observable neurons, the partial correlation computed from the observed neural signals is,
\begin{equation}
C_{ob}^{-1}=S_{ob} = S_{11}  -  S_{12}\cdot S_{22}^{-1}\cdot S_{21}
\end{equation}
because the invisible latent neurons as shown in Eq.~\ref{passiveModel} introduce correlations into the measurable system.
We denote this correlation introduced from the latent inputs as
\begin{equation}
L= S_{12}\cdot S_{22}^{-1}\cdot S_{21}
\end{equation}
If we can make the assumption that the connection between the visible neurons are sparse, i.e. $S_{11}$ is sparse and the number of latent neurons is much smaller than the number of visible neurons, i.e. $d << p$. Then, prior works \citep{chandrasekaran2011rank} have shown that if $S_{ob}$ is known, $S_{11}$ is sparse enough and L's rank is low enough (within the bound defined in \cite{chandrasekaran2011rank}), then the solution of 
\begin{equation}
S_{11}-L=S_{ob}
\end{equation}
 is uniquely defined and can be solved by the following convex optimization problem
\begin{equation}
\argmin\limits_{S_{11},L} ||S_{11}||_1 + \alpha*tr(L)
\end{equation}
under the constraint that 
\begin{equation}
S_{ob}=S_{11}-L
\end{equation}
Here, $||\ ||_1$ is the L1-norm of a matrix, and $tr()$ is the trace of a matrix. $\alpha$ is the penalty ratio between the L1-norm of $S_{11}$ and the trace of L and is set to $1/\sqrt{N}$ for all our estimations.
%Prior works \citep{chandrasekaran2011rank} shown that when $\alpha$ is within certain range, this convex problem always converge to the same solution.

However, the above method is used to regularize precision matrix. For our  differential covariance estimation, we need to make small changes to the derivation. Note that
if we assume the neural signals of the latent neurons are known, and let $l$ be the indexes of these latent neurons,
% \[L= \left( \begin{array}{c}
%latent_1\\
%latent_2\\
%...\\
%latent_d
%\end{array} \right)\] 
then from our previous   section (section~\ref{partial_diffCov}),
\begin{equation}
\Delta_{S_{i,j}} = \Delta_{P_{i,j}} - COV_{j,l}\cdot COV_{l,l}^{-1}\cdot \Delta_{C_{i,l}}^T
\end{equation}
removes the $V_{latent}$ terms in Eq.~\ref{passiveModel}. 

Even if $l$ is unknown, 
$$COV_{j,l}\cdot COV_{l,l}^{-1}\cdot \Delta_{C_{i,l}}^T$$
is low-rank, because it is bounded by the dimensionality of $COV_{l,l}$, which is $d$. And $\Delta_S$ is the internal connections between the visible neurons, which should be a sparse matrix.
Therefore, letting 
\begin{equation}
\begin{array}{c}
S_{ob}=\Delta_P \\
S_{11}=\Delta_S \\
L= - COV_{j,l}\cdot COV_{l,l}^{-1}\cdot \Delta_{C_{i,l}}^T
\end{array}
\end{equation}
we can use the original sparse+latent method to solve for $\Delta_S$. In this paper, we used the inexact robust PCA algorithm ( \url{http://perception.csl.illinois.edu/matrix-rank/sample_code.html } ) to solve this problem \citep{lin2011linearized}.

\subsection{the generalized linear model method}
\label{appGLM}

As summarized by \cite{roudi2015multi}, GLMs assume that every neuron spikes at a time-varying
rate which depends on earlier spikes (both those of other
neurons and its own) and on ‘external covariates’ (such as
a stimulus or other quantities measured in the experiment). 
As they explained, the influence of earlier spikes on the firing
probability at a given time is assumed to depend on
the time since they occurred. For each ‘pre-postsynaptic’
pair i, j, it is described by a function $J_{ij}(\tau)$ of this time lag \citep{roudi2015multi}.
In this paper, we average this temporal dependent function $J_{ij}(\tau)$ over time to obtain the functional connectivity estimation of this method.

The Spike trains used for the GLM method were computed using the spike detection algorithm from \cite{quiroga2004unsupervised} (\url{https://github.com/csn-le/wave_clus}). The paper's default parameter set is used except that the maximum detection threshold is increased. The code is provided in our github repository (\url{https://github.com/tigerwlin/diffCov/blob/master/spDetect.m}) for reference. The GLM code was obtained from \cite{pillow2008spatio} (\url{http://pillowlab.princeton.edu/code_GLM.html}) and applied on the spike trains.

\section{Details about the thalamocortical model}
\label{thalamo}
\subsection{Intrinsic currents}

For the thalamocortical model, a conductance-based formulation was used for all neurons. The cortical
neuron consisted of two compartments: dendritic and axo-somatic compartments,
similar to previous studies \citep{bazhenov02-8691, chen12-3987, bonjean11-9124} and is described by the following equations,
\begin{equation}
\begin{array}{rcl}
%C_m\frac{dV_D}{dt} & = &-ACh_{CX}I_d^{K-leak} -I^{leak}_d -I^{Na}_d -I^{Nap}_d -I^{Ca}_d -I^{Km}_d -I^{syn} \\
C_m\frac{dV_D}{dt} & = &-I_d^{K-leak} -I^{leak}_d -I^{Na}_d -I^{Nap}_d -I^{Ca}_d -I^{Km}_d -I^{syn} \\
g_c^{s}(V_S-V_D) & = & -I^{Na}_S-I^{K}_S-I^{Nap}_S
\end{array}
\end{equation}
where the subscripts s and d correspond to axo-somatic and dendritic
compartment,
% ACh$_{CX}$ is variable which is inversely proportional to the level
%of ACh in cortex ( ACh$_{CX}$ = 1 for awake, ACh$_{CX}$ = 1.25 for S2, ACh$_{CX}$ =  2.0 for N3
%and ACh$_{CX}$=0.85 during REM stage) and determines the potassium leak current
%(I$^{K-leak}$), 
I$^{leak}$ is the Cl$^-$ leak currents, I$^{Na}$ is fast Na$^+$ channels, I$^{Nap}$ is persistent sodium current,
I${^K}$ is fast delayed rectifier K$^+$ current, I$^{Km}$ is slow voltage-dependent non-inactivating K$^+$
current, I$^{KCa}$ is slow Ca$^{2+}$ dependent K$^{+}$ current, I$^{Ca}$ is high-threshold Ca$^{2+}$ current, I$^{h}$ is
hyperpolarization-activated depolarizing current and I$^{syn}$ is the sum of
synaptic currents to the neuron. All intrinsic currents were of the
form: $g(V-E)$, where g is the conductance, V is the voltage of the corresponding
compartment and E is the reversal potential. The detailed descriptions
of individual currents are provided in previous publications \citep{bazhenov02-8691, chen12-3987}. The
conductance of the leak currents were 0.007 mS/cm$^2$ for I$_{d}^{K-leak}$ and 0.023 mS/cm$^2$
for I$_{d}^{leak}$. The maximal conductance for different currents were, I$_{d}^{Nap}$: 2.0 mS/cm$^2$,
I$_{d}^{Na}$: 0.8 mS/cm$^2$, I$_{d}^{Km}$: 0.012 mS/cm$^2$, I$_{d}^{KCa}$: 0.015 mS/cm$^2$ , I$_{d}^{Km}$: 0.012 mS/cm$^2$, I$_{s}^{Na}$: 3000mS/cm$^2$, I$_{s}^{K}$: 200 mS/cm$^2$ and I$_{s}^{Nap}$: 15 mS/cm$^2$. C$_{m}$ was $\SI{0.075}{\micro\farad\per\centi\meter^2}$.

The following describes the IN neurons:
\begin{equation}
\begin{array}{rcl}
%C_m\frac{dV_D}{dt} & = &-ACh_{CX}I_d^{K-leak} -I^{leak}_d -I^{Na}_d -I^{Ca}_d -I^{KCa}_d -I^{Km}_d -I^{syn} \\
C_m\frac{dV_D}{dt} & = &-I_d^{K-leak} -I^{leak}_d -I^{Na}_d -I^{Ca}_d -I^{KCa}_d -I^{Km}_d -I^{syn} \\
g_c^{s}(V_S-V_D) & = & -I^{Na}_S-I^{K}_S
\end{array}
\end{equation}
The conductance for leak currents for IN neurons were 0.034 mS/cm$^2$
for I$_{d}^{K-leak}$ and 0.006 mS/cm$^2$ for I$_{d}^{leak}$. Maximal conductance for other currents were, I$_{d}^{Na}$ :
0.8 mS/cm$^2$, I$_{d}^{Ca}$: 0.012 mS/cm$^2$, I$_{d}^{KCa}$: 0.015 mS/cm$^2$ , I$_{d}^{Km}$: 0.012 mS/cm$^2$, I$_{s}^{Na}$: 2500
mS/cm$^2$ and I$_{s}^{K}$: 200 mS/cm$^2$ .

The TC neurons consisted of only single compartment and was described
as follows,
\begin{equation}
\begin{array}{rcl}
%\frac{dV_D}{dt} & = &-ACh_{TC}I^{K-leak} -I^{leak} -I^{Na} -I^{K} -I^{LCa} -I^{h} -I^{syn} \\
\frac{dV_D}{dt} & = &-I^{K-leak} -I^{leak} -I^{Na} -I^{K} -I^{LCa} -I^{h} -I^{syn} \\
\end{array}
\end{equation}
%
%where ACh$_{TC}$ was 1, 1.25, 0.85 during awake, S2, N3 and REM stage
%correspondingly. 
The conductance of leak currents were, I$^{leak}$: 0.01 mS/cm$^2$,
I$^{K-leak}$: 0.007 mS/cm$^2$. The maximal conductance for other currents were, fast
Na$^+$ (I$^{Na}$) current: 90 mS/cm$^2$, fast K$^+$ (I$^{k}$) current: 10 mS/cm$^2$, low threshold
Ca$^{2+}$(I$^{LCa}$) current: 2.5 mS/cm$^2$, hyperpolarization-activated depolarizing
current (I$^{h}$): 0.015 mS/cm$^2$. 
%The effect of HA on I$^{h}$ was implemented as a
%shift in its activation function based on experimental evidences as
%follows, 
%\begin{equation}
%\begin{array}{rcl}
%I^{h} & = &g([O] + k[O_L])(V-E_h) \\
%[O] = ......
%k1 = h_\inf/\tau_s
%k2 = (1-h_\inf)/\tau_s
%h_\inf=frac{1}{1+exp(v+75-Shift_{HA})/5.5}
%\end{array}
%\end{equation}
%where Shift$_{HA}$ was -12mV, -2.5mV, 0mV and 2.0mV for awake, S2, N3
%and REM stage correspondingly. The values of other constants were
%same as previous study \citep{bonjean11-9124}. 

The RE cells were also modeled as a single compartment neuron as follows,
\begin{equation}
\begin{array}{rcl}
%\frac{dV_D}{dt} & = &-ACh_{TC}I^{K-leak} -I^{leak} -I^{Na} -I^{K} -I^{LCa} -I^{h} -I^{syn} \\
\frac{dV_D}{dt} & = &-I^{K-leak} -I^{leak} -I^{Na} -I^{K} -I^{LCa} -I^{h} -I^{syn} \\
\end{array}
\end{equation}
%
%where ACh$_{RE}$ was 1, 0.8, 0.5 and 1.15 for awake, S2, N3 and REM
%stage. Since it is known that ACh reduces excitability of the RE cells
%(citation), the ACh$_{RE}$ is reduced during NREM stage compared to awake. 
The conductance
for leak currents were, I$^{leak}$: 0.05 mS/cm$^2$, I$^{K-leak}$: 0.016 mS/cm$^2$. The maximal
conductance for other currents were, fast Na$^+$ (I$^{Na}$) current: 100 mS/cm$^2$,
fast K$^+$ (I$^{K}$) current: 10 mS/cm$^2$, low threshold Ca$^{2+}$(I$^{LCa}$) current: 2.2 mS/cm$^2$.

\subsection{Synaptic currents}

GABA-A, NMDA and AMPA synaptic currents were described by first-order
activation schemes \citep{timofeev00-1185}. The equations for all synaptic currents used
in this model are given in our previous publications \citep{bazhenov02-8691, chen12-3987}. Briefly,
below we mention only the relevant equations.
% used to model the
%effect of levels of ACh and GABA.
%
\begin{equation}
\begin{array}{rcl}
%I_{syn}^{AMPA}$=$L_{ACh}g_{syn}[O](V-E_{AMPA}) \\
%I_{syn}^{GABA}$=$L_{GABA}g_{syn}[O](V-E_{GABA})
I_{syn}^{AMPA}$=$g_{syn}[O](V-E_{AMPA}) \\
I_{syn}^{NMDA}$=$g_{syn}[O](V-E_{NMDA}) \\
I_{syn}^{GABA}$=$g_{syn}[O](V-E_{GABA})
\end{array}
\end{equation}
%
%L$_{ACh}$ was 1, 1.25, 2.0 and 0.8 for awake, S2, N3 and REM stages
%for cortical PY-PY, TC-PY and TC-IN connections. L$_{GABA}$ was 1, 1.25, 2.0
%and 0.55 for awake, S2, N3 and REM stages for cortical IN-PY,
%RE-RE and RE-TC connections. 

\section{Supplementary figures}
\label{suppFig}

\begin{figure}[hbtp]
\begin{center}
\includegraphics[width=\textwidth]{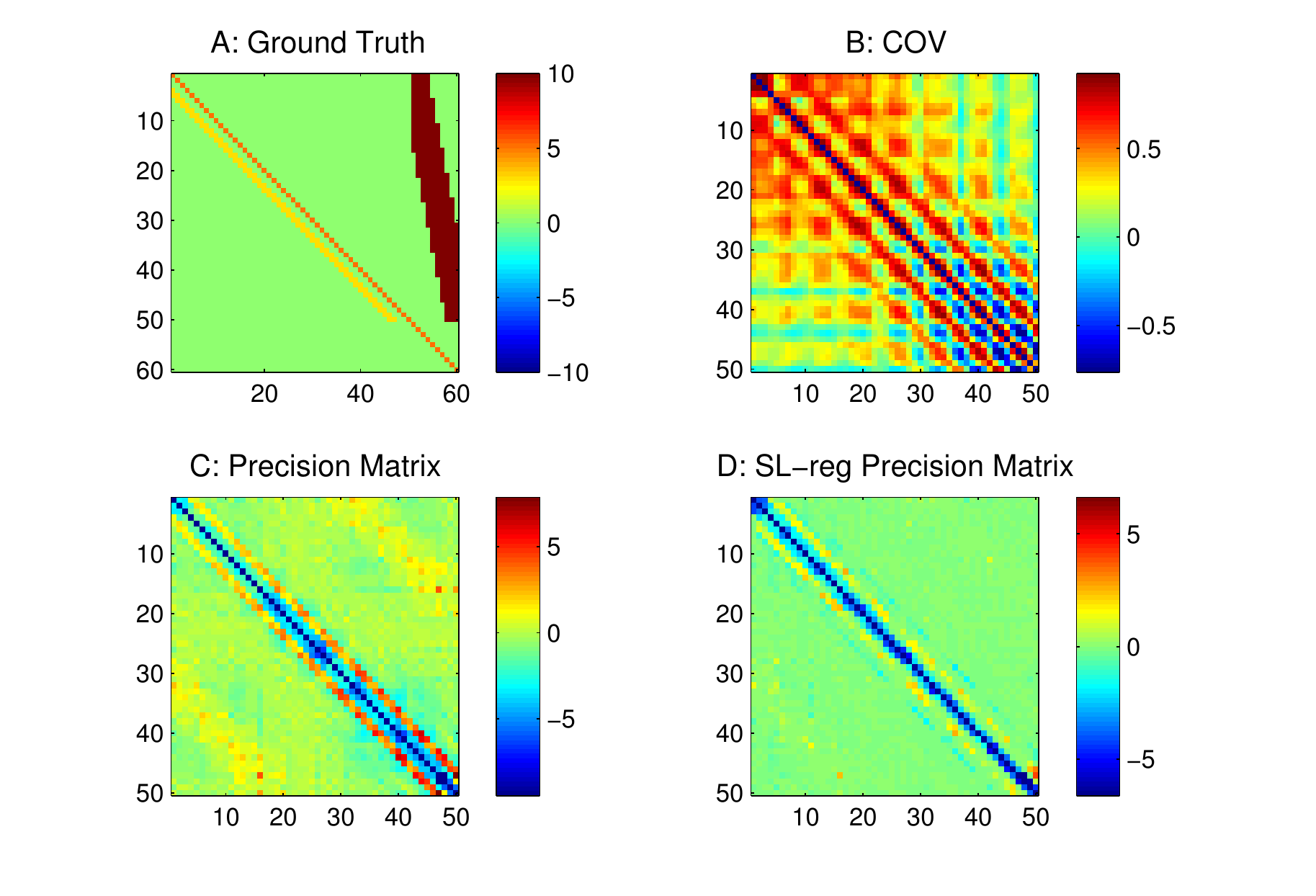}
\end{center}
\caption{Supplementary1, passive neuron model with 5 ms fixed synaptic delay. Results from correlation-based methods. A) Ground truth connection matrix. neurons 1-50 are visible neurons. neurons 51-60 are invisible neurons. B) Estimation from the correlation method. C) Estimation from the precision matrix. D) Sparse+latent regularized precision matrix.}
\label{Supplementary1}
\end{figure}
\begin{figure}[hbtp]
\begin{center}
\includegraphics[width=\textwidth]{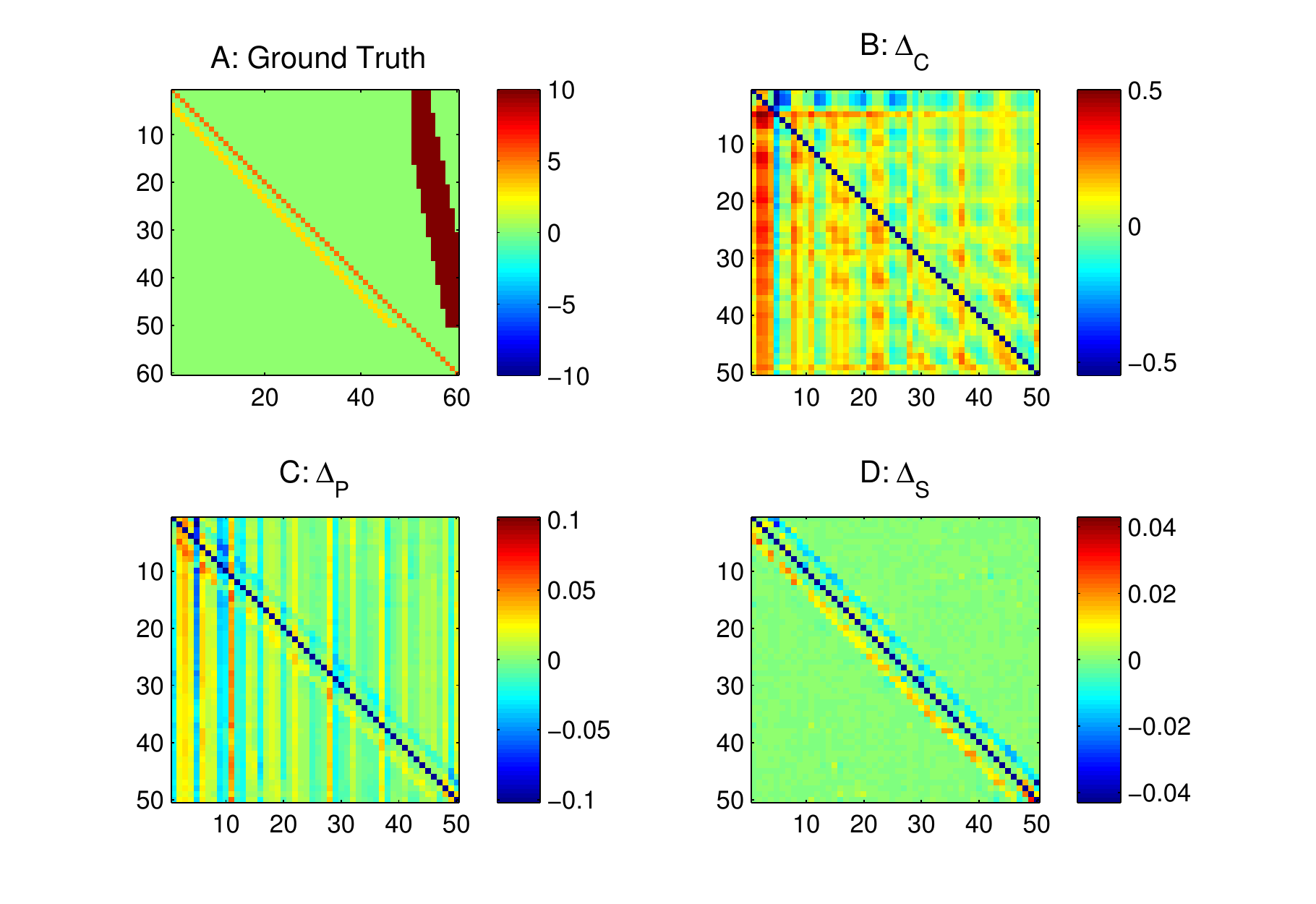}
\end{center}
\caption{Supplementary2, differential covariance analysis of the passive neuron model with 5 ms fixed synaptic delay. The color in B,C,D indicates direction of the connections. For element $A_{ij}$, warm color indicates $i$ is the sink, $j$ is the source, i.e. $i \leftarrow j$, and cool color indicates $j$ is the sink, $i$ is the source,  i.e. $i \rightarrow j$. A) Ground truth connection matrix. B) Estimation from the differential covariance method. C) Estimation from the  partial differential covariance method. D) Estimation from the sparse+latent regularized partial differential covariance method.}
\label{supplementary2}
\end{figure}

\begin{figure}[hbtp]
\begin{center}
\includegraphics[width=\textwidth]{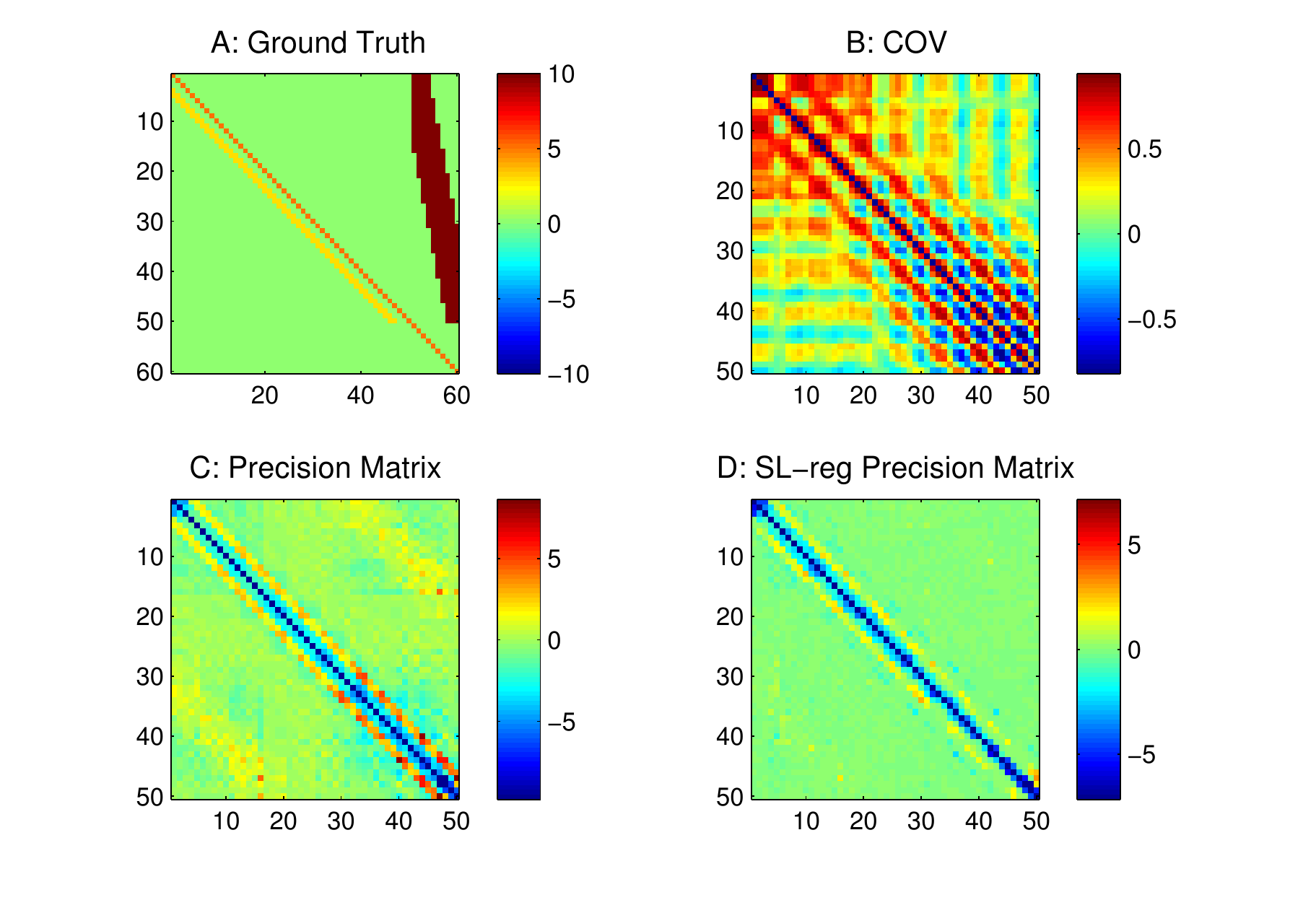}
\end{center}
\caption{Supplementary3, passive neuron model with 0-10 ms uniformly distributed synaptic delay. Results from correlation-based methods. A) Ground truth connection matrix. neurons 1-50 are visible neurons. neurons 51-60 are invisible neurons. B) Estimation from the correlation method. C) Estimation from the precision matrix. D) Sparse+latent regularized precision matrix.}
\label{Supplementary3}
\end{figure}

\begin{figure}[hbtp]
\begin{center}
\includegraphics[width=\textwidth]{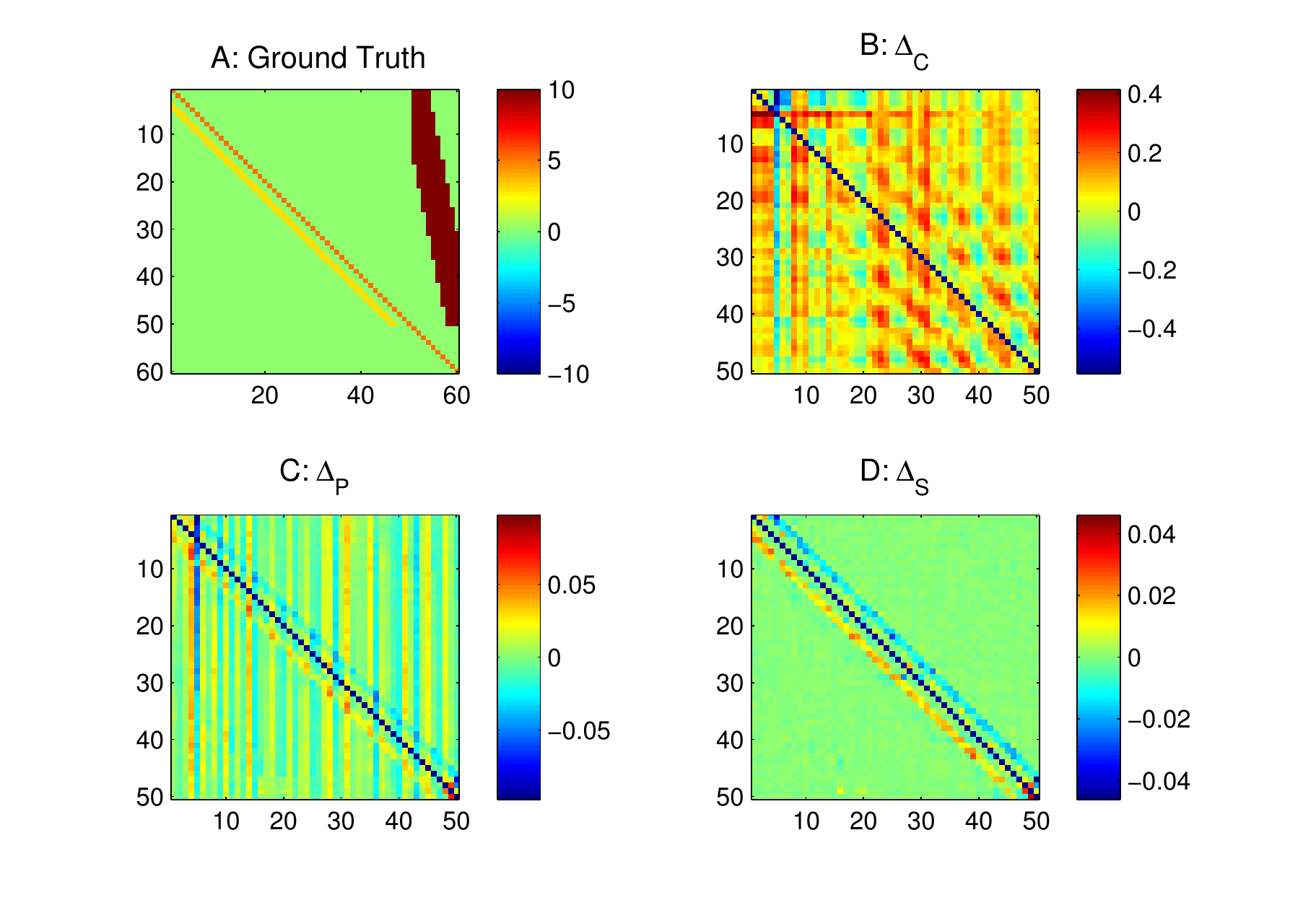}
\end{center}
\caption{Supplementary4, differential covariance analysis of the passive neuron model with 0-10 ms uniformly distributed synaptic delay. The color in B,C,D indicates direction of the connections. For element $A_{ij}$, warm color indicates $i$ is the sink, $j$ is the source, i.e. $i \leftarrow j$, and cool color indicates $j$ is the sink, $i$ is the source,  i.e. $i \rightarrow j$. A) Ground truth connection matrix. B) Estimation from the differential covariance method. C) Estimation from the  partial differential covariance method. D) Estimation from the sparse+latent regularized partial differential covariance method.}
\label{supplementary4}
\end{figure}

\pagebreak
\bibliographystyle{apalike}
\bibliography{ref}

\end{document}